\documentclass[superscriptaddress,showkeys,useAMS,twocolumn]{revtex4-1}
\usepackage{graphics, graphicx, amsmath, amssymb}
\usepackage{epstopdf}
\usepackage{dsfont}
\usepackage[dvipsnames]{color}
\usepackage{ulem}
\begin{document}

\title{Emergence of singularities from decoherence: quantum catastrophes}
\date{\today}
\author{Aaron Z. Goldberg}
\affiliation{Department of Physics and Astronomy, McMaster University, 1280 Main Street W., Hamilton, Ontario, Canada, L8S 4M1}
\affiliation{Department of Physics, University of Toronto, 60 St. George Street, Toronto, Ontario M5S 1A7, Canada}
\author{Asma Al-Qasimi}
\affiliation{Department of Physics and Astronomy, University of Rochester, Rochester, New York 14627, USA.}
\author{J. Mumford}
\affiliation{Department of Physics and Astronomy, McMaster University, 1280 Main Street W., Hamilton, Ontario, Canada, L8S 4M1}
\affiliation{School of Arts and Sciences, Red Deer College, 100 College Boulevard, Red Deer, Alberta, T4N 5H5, Canada}
\author{D. H. J. O'Dell}
\affiliation{Department of Physics and Astronomy, McMaster University, 1280 Main Street W., Hamilton, Ontario, Canada, L8S 4M1}

\begin{abstract}
{We use a master equation to study the dynamics of two coupled macroscopic quantum systems (e.g.\ a Josephson junction made of two Bose-Einstein condensates or two spin states of an ensemble of trapped ions) subject to a weak continuous measurement.  If the coupling between the two systems is suddenly switched on the resulting dynamics
leads to caustics (fold and cusp catastrophes) in the number-difference probability distribution, and at the same time the measurement gradually induces a quantum-to-classical transition. Decoherence is often invoked to help resolve paradoxes associated with macroscopic quantum mechanics, but here, on the contrary, caustics are well-behaved in the quantum (many-particle) theory and divergent in the classical (mean-field) theory. Caustics thus represent a breakdown of the classical theory towards which decoherence seems to inevitably lead. We find that measurement backaction plays a crucial role in softening the resulting singularities and calculate the modification to the Arnol'd  index which governs the scaling of the caustic's amplitude with the number of atoms. The Arnol'd index acts as a critical exponent for the formation of singularities during quantum dynamics and its modification by the open nature of the system is analogous to the modification of the critical exponents of phase transitions occurring in open systems.  }
\end{abstract}
\maketitle

\section{Introduction}

Decoherence plays an important role in our current understanding of the quantum-to-classical transition. In particular, it provides an explanation for why macroscopic quantum superpositions (Schr\"{o}dinger cat states) are not observed in everyday life: decoherence destroys quantum interference at a rate that depends sensitively upon the size of the superposition, and thereby reduces the associated probability distribution to a sum of classical probabilities \cite{Zurek1991,Zurek2003,Schlosshauer2005}. This occurs whenever a system becomes entangled with its environment, causing a continuously monitored quantum system to behave classically \cite{Joos85,Zurek1993b,Zurek94,Gallis96}, and even to display elements of chaotic dynamics (something 
that is generally absent in quantum mechanics) \cite{Spiller94,Habib98,Bhattacharya2000, Karkuszewski2002}.  Pioneering experiments using atoms interacting with electromagnetic fields  have observed the decay to classicality  \cite{Brune1996,Myatt2000,Turchette2000,Kokorowski2001,Hornberger2003,Deleglise2008}, including  the appearance of chaos \cite{Steck2001,Hensinger2001,Steck2002,Chaudhury2009}, and measurement-induced suppression of tunneling \cite{Patil2015}.

In this paper we consider the measurement-induced decoherence of a pair of coupled macroscopic quantum systems. For concreteness we will base our discussion on a bosonic Josephson junction (BJJ) made from two atomic Bose-Einstein condensates (BECs) coupled via a tunneling barrier \cite{Java86}. These have been demonstrated in a number of experiments \cite{Albiez2005,Schumm2005,Levy2007,Zibold10,Leblanc11,Trenkwalder16} where the interatomic interactions introduce a nonlinearity responsible for macroscopic quantum self-trapping \cite{Albiez2005,Levy2007,Milburn1997,Smerzi1997,Raghavan1999} and a symmetry breaking phase transition \cite{Trenkwalder16}. Both these phenomena can be understood at the classical-field level of the Gross-Pitaevskii equation (GPE) \cite{Pitaevskii2001}, and at the many-body level the interactions generate entanglement which manifests itself as reduced atom number fluctuations \cite{Orzel2001,Esteve2008,Gross2010} (and increased phase fluctuations) between the BECs. This entanglement has been characterized in terms of Fisher information \cite{Strobel2014} and used to perform sub-shot noise magnetometry \cite{Muessel2014}. However, the results we find in this paper also apply to other macroscopic quantum systems such as trapped ions with two spin states  which can be used, for example, to realize the transverse field Ising model (TFIM) with long-range interactions  \cite{Britton12,Jurcevic14,Richerme14,Bohnet16} and the Dicke model \cite{Safavi18}, or, indeed, solid state realizations of Josephson junctions such as SQUIDs  \cite{Everitt2009,Devoret2013}.

A macroscopic description of a BJJ can be given in terms of the conjugate variables $\phi=\phi_{L}-\phi_{R}$ and $z = (N_{L}-N_{R})/N$ \cite{Smerzi1997,Raghavan1999,Leggett01} which are the phase and relative number difference, respectively, between the left- and right-hand BECs, and $N$ is the total number of atoms. 
$\phi$ can be measured via a matter-wave version of Young's double slit experiment in which the BECs are released from their trapping potential to produce an interference pattern \cite{Andrews97,Cirac1996,Castin1997,Sinatra98}, and this technique has been used experimentally to study dephasing \cite{Hofferberth07,Gring2012,AduSmith2013,Langen2013}. Nondestructive phase measurements can be performed via light scattering \textit{in situ} \cite{Saba2005}. Instead of measuring $\phi$, the number difference $z$ can be measured by light absorption imaging \cite{Albiez2005} or phase contrast \cite{Levy2007} imaging, where the latter method can also be nondestructive.

There have been a number of previous theoretical studies of a BJJ subject to decoherence \cite{Ruostekoski1998,Huang06,Khodorkovsky08,Witthaut09,Ferrini2010,Javanainen2013}. The particular case we study here, a weak continuous measurement of $z$, has been studied in \cite{Javanainen2013} where it was found that if the measurements are frequent enough to resolve the dynamics then measurement backaction causes the system to behave classically. In this context, `classical' means `mean-field' indicating the absence of many-body entanglement; in coordinate space the condensates still obey the Gross-Pitaevskii wave equation, which is a classical field equation conceptually analogous to Maxwell's wave equation for light in the sense that neither take account of the quantization of field excitations (atoms and photons, respectively). 

The specific dynamics we consider here arise from a quench in the tunneling rate from zero to a finite value, describing the case of two independent condensates suddenly placed in contact \cite{Zapata2003,Xiong2006,Trujillo2009}. This sets the combined system into motion and a series of collapses and partial revivals of the many-body state occur as a function of time \cite{Milburn1997,Veksler2015}.  The revivals can be interpreted as caustics  akin to those studied in optics, but living in the Fock space inhabited by the number-difference amplitudes rather than coordinate space \cite{Odell2012,Mumford2017,Mumford2019}. 

Caustics in optics are singular in the classical limit (ray optics). Specifically, caustics are regions of infinite intensity indicating a fundamental breakdown of the ray theory. The singularity can be tamed by going up one step in complexity to the wave theory and thereby including interference which is the crucial ingredient needed to smooth a ray caustic. Interference turns the divergent ray caustic into a smooth and non-divergent interference pattern known as a ``wave catastrophe'' \cite{Berry1981}. However, in our case the classical description is via the GPE which is already a wave equation and hence includes wave interference in coordinate space. Nevertheless, the dynamics of the GPE does generate singularities but this time in Fock space, which is the natural arena where many-body dynamics takes place. Such singularities are examples of ``quantum catastrophes''   and represent a breakdown of classical field theory \cite{Leonhardt2002,Berry2004,Berry2008}. Analogously to how ray caustics are regulated by the inclusion of phase interference, quantum catastrophes can be regulated by going up another step in complexity to the many-body theory where the discreteness of the quanta  (in our case atom number difference) regularizes the caustic \cite{Odell2012,Mumford2017,Mumford2019}.

Whereas the inclusion of decoherence usually helps reconcile macroscopic quantum mechanics with our everyday experience, we now see that in the presence of caustics it has the potential to instead lead back to the paradoxical situation of a theory with singularities. However, the missing ingredient in this description is how the quantum noise introduced by the measurement backaction affects the catastrophe, and this plays an important role in our investigation.

In a recent paper by Naghiloo \textit{et al} \cite{Naghiloo17}, the dynamics of a continuously measured qubit have been analyzed, both theoretically and experimentally. Specifically, the measured fluorescence signal from a driven superconducting qubit was used to construct a quantum trajectory representing the time evolution of its quantum state during each experimental run. The random nature of quantum measurement outcomes means that these trajectories have the appearance of a stochastic or random walk and can be simulated using a stochastic master equation; summing over the trajectories leads to the same probability distribution as would be obtained using an ordinary (non-stochastic) master equation. Naghiloo \textit{et al} observed that caustic-like structures can emerge from among the ensemble of quantum trajectories: regions or paths of high probability in the configuration space of the qubit as a function of time. They used the term ``quantum caustics'' to describe these caustics because they are made from quantum trajectories as opposed to, say, ordinary optical caustics formed from geometrical light rays.


Our work in this paper is related to that of \cite{Naghiloo17}, but differs in some significant ways. One of the main differences is that we consider a many-particle system described by a discrete Fock space. The ``quantum caustics/catastrophes'' that we find are  quantum due to the discreteness of Fock space even without the effect of quantum measurements. The quantum probability distribution in Fock space is discrete but also displays interference fringes which, in the semiclassical regime $N>>1$, lead to structures resembling the wave catastrophes introduced above. Wave catastrophes obey a remarkable set of scaling relations determined by the Arnol'd and Berry indices  \cite{Berry1981}. We use an ordinary master equation to see the effect that weak continuous measurement has on these pre-existing caustics in the probability distribution; in this sense this paper is about decoherence of a quantum catastrophe. We find that for weak measurements and short times the caustics maintain their qualitative shapes although the measurement backaction changes them in subtle ways such as modifying the scaling exponents.

The rest of this paper is organized as follows: in Section \ref{sec:classical} we give the classical (GPE) description of the dynamics of a BJJ, and we follow this in Section \ref{sec:quantum} with the equivalent quantum (Bose-Hubbard model) description. In Section \ref{sec:master} we introduce the master equation describing a continuous measurement and in Section \ref{sec:decoherence} we present results showing the effects this has upon the caustics, specializing in Section \ref{sec:arnold} to showing how the Arnol'd scaling exponent for the amplitude of a wave catastrophe is modified by decoherence. We give our conclusions in Section \ref{sec:conclusions}. There are also six appendices where we provide background information on topics such as the phase space dynamics, structural stability of catastrophes as well as the details of various derivations and calculations.

\section{Classical dynamics}
\label{sec:classical}

\begin{figure*}\includegraphics[width=\textwidth]{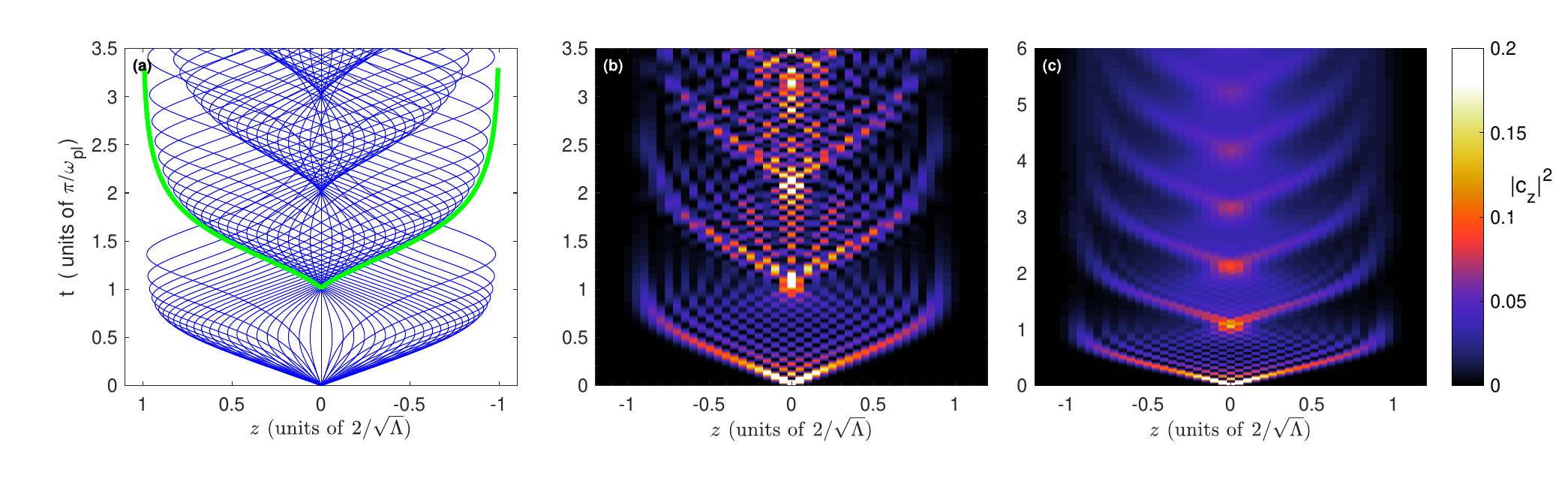}
	\caption{Numerical solutions of the dynamics following a quench in a BJJ showing classical and quantum catastrophes. \textbf{(a)} The classical trajectories for the fractional number difference $z$ periodically form classical cusp caustics. Each trajectory has a different value of the initial phase difference  ranging from $-\pi$ to $\pi$ in steps of $\pi/25$, which samples the initial quantum distribution in accordance with the TWA. The tilt bias $\Delta E=0$.
\textbf{(b)}	 The quantum probability distribution for the number difference ($|c_z|^2=\rho_{z,z}$) over time. The quantized probability distribution resembles certain features of the classical trajectory density but with discretized values of number difference giving rise to quantum cusp catastrophes. Interference fringes whose analytic form is described by a discrete Pearcey function \cite{Mumford2019} are visible within the arms of the caustics.
\textbf{(c)} Decoherence is turned on in the quantum evolution ($D=0.1$). For this value of $D$ the quantum interference remains visible for a few oscillations\textcolor{black}{,} but at later times this is washed out and even the caustics are diffuse. When $t \rightarrow \infty$ the system tends to $\rho_\mathrm{steady}$ [Eq. (\ref{eq:steady-state-eqn})].
$\Lambda=25$ for all panels, and $N=100$ for \textbf{(b-c)}.}
	\label{fig:Classical-trajectories-MF-normal-and-tilt-v2}
\end{figure*}

Catastrophe theory predicts that only certain shapes of singularity are structurally stable (stable against perturbations)  and hence occur frequently in nature with no requirement for special symmetry. In four or fewer dimensions these are Thom's famous seven elementary catastrophes, each of which forms an equivalence class \cite{Thom1975,Arnold1975}. Well-known everyday examples include rainbows (fold catastrophes) and the bright cusp shape (cusp catastrophe) formed in a coffee cup on a sunny day \cite{Berry1981}. More generally, catastrophes occur in hydrodynamics as rogue waves \cite{Hohmann10} and tidal bores \cite{Berry18}, and have also been observed in atom optics experiments  \cite{Rooijakkers03,Huckans09,Rosenblum14}. In the present quenched BJJ problem catastrophes appear as cusp caustics formed by the envelopes of families of classical trajectories in the $(z,t)$-plane, as shown in Fig.\ \ref{fig:Classical-trajectories-MF-normal-and-tilt-v2}(a) (for the phase space representation of the dynamics please see Appendix \ref{app:phasespace} and Fig.\ \ref{fig:phasespacedynampic}). Each trajectory is a solution of Josephson's equations \cite{Smerzi1997,Raghavan1999}  
\begin{equation}
\begin{aligned}
\frac{\partial z}{\partial \tau}=&-\frac{\partial h}{\partial \phi}=-\sqrt{1-z^2}\sin\phi\\
\frac{\partial \phi}{\partial \tau}=&\frac{\partial h}{\partial z}=\Lambda z+\frac{z}{\sqrt{1-z^2}}\cos\phi+\Delta E ,
\end{aligned}
\label{eq:Hamilton's equations}
\end{equation}
which correspond to the GPE in the number-phase representation in the tight-binding regime  \cite{Pitaevskii2001}. As indicated, they can be obtained as Hamilton's equations of motion from the dimensionless Hamiltonian $h=H/E_{J}$ where 
\begin{equation}
H=\frac{E_C}{2}\left(\frac{Nz}{2}\right)^2-E_J\sqrt{1-z^2}\cos\phi+E_J\, \Delta E \, z .
\label{eq:mean-field-H}
\end{equation}
Here, $E_C$ characterizes the interatomic interactions, $E_J$ is the coupling energy between the two BECs, $\Delta E$ is a tilt bias (if present) in units of the coupling energy, and time is scaled as $\tau\equiv t\left(2E_J/N\hbar\right)$. Thus, the classical dynamics are governed by just two parameters: $\Lambda \equiv E_CN^2/4E_J$ and $\Delta E$. In the fully quantum (many-body) description we need to additionally specify $N$ which plays a role analogous to $1/\hbar$ \cite{Vardi2001}.

\begin{figure*}
	\includegraphics[width=\textwidth]{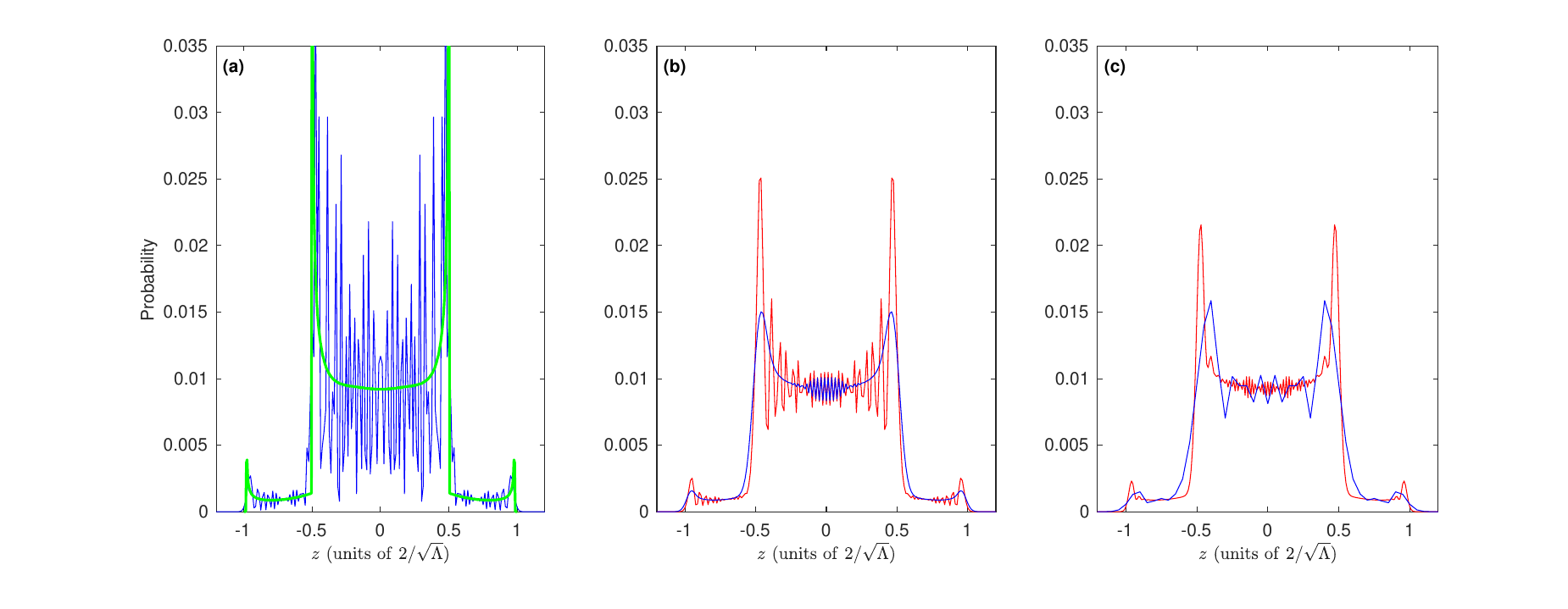}
	\caption{Comparison of the  probability distributions for the fractional number difference $z$ for classical, quantum, and quantum with decoherence dynamics in a BJJ at the time slice $t=1.5\pi / \omega_\mathrm{pl}$ with $\Lambda=25$. \textbf{(a)} Classical (thick green line) versus the quantum with no decoherence (thin blue line), with the latter having $N=400$ particles. The classical probability diverges as the inverse square root of the distance from the fold caustics at 
$z \approx \pm 0.5$ and $z \approx \pm 1$ (in units of $2/\sqrt{\Lambda}$). The quantum probability distribution, which is always finite, is actually discrete in $z$ but drawn here with a continuous line. \textbf{(b)} Both distributions are for the quantum case with $N=400$ but with different decoherence strengths: $D=0.025$ (red) and $D=0.4$ (blue).  \textbf{(c)} Both distributions are for the quantum case with $D=0.1$ but with different numbers of particles:  $N=100$ (blue) and $N=500$ (red). }
	\label{fig:time slice comparing classical quantized decoherence3}
\end{figure*}

The method of using an ensemble of non-interfering classical trajectories to approximate the quantum dynamics is sometimes referred to as the \textit{truncated Wigner approximation} (TWA) \cite{Javanainen2013,Sinatra2002,Polkovnikov2003}. The idea is to choose multiple initial conditions for the dynamical variables $(\phi,z)$ by sampling them from a probability distribution obtained from the full quantum theory but evolve these initial values using the classical equations of motion, which in our case are the Josephson equations given in Eq.\ (\ref{eq:Hamilton's equations}). Summing over the time-evolved classical trajectories thereby includes the quantum noise present in the initial quantum wave function.

In the present case our initial state consists of two independent BECs. This implies that there is no initial coherence between the two BECs and also that the number difference $z$ is a precise number which we take to be zero. The latter condition  means that the two BECs have exactly the same number of atoms, but by virtue of their structural stability, the caustics we obtain are qualitatively unaffected if instead we take $z \neq 0$ (we explore the structural stability of caustics further in  Appendix \ref{app:stability to initial conditions}.) In the classical theory the conjugate variables $z$ and $\phi$ are continuous and commute with each other, whereas in the quantum theory they obey the commutator $[\hat{\phi},\hat{z}] \approx 2 i/N$ \cite{Paraoanu2001}, at least when $N$ is moderately large.  Thus, Heisenberg's uncertainty relation implies that the initial phase difference $\phi$ must be completely undefined.  The initial values we use for the classical trajectories are therefore $z=0$ for all, but $\phi$ uniformly distributed over the range $-\pi$ to $\pi$. The resulting classical trajectories are those plotted in Fig.\ (\ref{fig:Classical-trajectories-MF-normal-and-tilt-v2})(a). 

According to catastrophe theory, the stable singularities in two dimensions [the $(z,t)$-plane] are fold lines that meet at cusp points \cite{Berry1981}. This is precisely what we see in Fig.\ \ref{fig:Classical-trajectories-MF-normal-and-tilt-v2}(a) which presents the results of solving Josephson's coupled equations numerically (note that energy conservation implies that all trajectories lie within the range $ -2/\sqrt{\Lambda} \le z \le 2/\sqrt{\Lambda}$). The thick green line traces the first cusp caustic, and thereafter cusp points occur periodically at the plasma frequency $\omega_\mathrm{pl}=\sqrt{2 E_J \left(2 E_J +N^2 E_c/2 \right)}/N\hbar=\left(2E_J/N\hbar\right)\sqrt{1+\Lambda}$, corresponding, in the harmonic approximation, to the frequency of motion around the bottom of the cosine potential well in the Hamiltonian.  
The density of trajectories diverges on a caustic, as shown by the thick green line in Fig.\ \ref{fig:time slice comparing classical quantized decoherence3}(a),  which plots the probability distribution at the time $1.5 \pi /\omega_{\mathrm{pl}}$ halfway between the first and second cusp points.

These classical predictions are equivalent to solving Liouville's equation $\partial \rho_{\mathrm{cl}} / \partial t=-\{\rho_{\mathrm{cl}} ,H\}$ for the classical phase space density $\rho_{\mathrm{cl}}(\phi,z)$, where $\{ \cdot \ , \ \cdot \}$ denotes the Poisson bracket. The dynamics in phase space are discussed in Appendix \ref{app:phasespace}, and snapshots at different times are presented in Fig.\ \ref{fig:phasespacedynampic} where $\rho_{\mathrm{cl}}$ is given by the density of points. The phase space density corresponding to the present initial conditions corresponds to a horizontal line along $z=0$; the dynamics winds this line into a whorl. The classical probability distribution for $z$ alone is obtained by projecting $\rho_{\mathrm{cl}}$ onto the $z$-axis by performing the integral $\int_{-\pi}^{\pi} \rho_{\mathrm{cl}}(\phi,z) \, \mathrm{d} \phi$. This procedure gives the thick green curve in Fig.\ \ref{fig:time slice comparing classical quantized decoherence3}(a) for the particular time $t=1.5\pi / \omega_\mathrm{pl}$. In this way one sees that the caustics in the probability distribution for $z$ arise from the horizontal portions of the whorl. We note that phase space whorls have been previously studied  for the case of the rigid pendulum in Reference \cite{BerryODell1999}; in that case it is possible to obtain analytic expressions for $\rho_{\mathrm{cl}}(\phi,z)$ in the long-time limit.

\section{Quantum dynamics}
\label{sec:quantum}

A fully quantum description of the dynamics can be achieved with the Bose-Hubbard model. A crucial difference between this and the classical treatment is that the number difference becomes quantized, recognizing the discrete nature of field quanta. We consider two sites (L and R) occupied by bosons that are created and annihilated by the operators $\hat{a}^\dagger_L$ ($\hat{a}^\dagger_R$) and $\hat{a}_L$ ($\hat{a}_R$), obeying bosonic commutation relations $[\hat{a}_{i},\hat{a}^{\dag}_{j}]=\delta_{ij}$. In terms of these operators, the fractional number-difference operator is defined as $\hat{z}\equiv (\hat{a}_L^\dagger\hat{a}_L-\hat{a}_R^\dagger\hat{a}_R )/N$. The problem can mapped onto a spin model by using the Schwinger mapping \cite{Sakurai2011} to angular momentum operators $\hat{J}_x \equiv (\hat{a}_L^\dagger\hat{a}_R+\hat{a}_R^\dagger\hat{a}_L)/2$, $\hat{J}_y \equiv i(\hat{a}_R^\dagger\hat{a}_L-\hat{a}_L^\dagger\hat{a}_R)/2$, and $\hat{J}_z \equiv (\hat{a}_L^\dagger\hat{a}_L-\hat{a}_R^\dagger\hat{a}_R)/2$, whence the Hamiltonian becomes \cite{Hines2003}
\begin{eqnarray}
\frac{\hat{H}}{2E_J/N} & = &\frac{N\Lambda}{4}\hat{z}^2-\frac{1}{2}\left(\hat{a}_L^\dagger\hat{a}_R+\hat{a}_R^\dagger\hat{a}_L\right)  \nonumber \\ & = & \frac{\Lambda}{N}\hat{J}_z^2-\hat{J}_x 
\label{eq:Hamiltonian-with-angular-momentum}
\end{eqnarray}
which describes a collective spin of total length $N/2$ made up of $N$ elementary spin-1/2 particles and is a special case of the Lipkin-Meshkov-Glick model  \cite{Zibold10,Lipkin65}. One point to notice with Eq.\ (\ref{eq:Hamiltonian-with-angular-momentum}) is its scaling with $N$, which is most clearly seen on the righthand side.  The maximum values that $\langle \hat{J}_{x} \rangle$ and $\langle \hat{J}_{z} \rangle$ can take are $N/2$,  and so the factor of $1/N$ multiplying the $\hat{J}_{z}^2$ term not only ensures that the Hamiltonian is extensive but also that both terms remain relevant in the thermodynamic limit $N \rightarrow \infty$.

In the absence of decoherence, the state of the system can be represented by the pure state  $\left|\Psi(t)\right\rangle=\sum_{z}c_{z}(t) \left|z\right\rangle$, where $\left|z\right\rangle$ is the number-difference (Fock) basis. Taking the initial state to be $\left|z=0\right \rangle$ (see Fig.\  \ref{fig:initial-gaussians} in Appendix \ref{app:stability to initial conditions} for the generalization of the initial condition to a gaussian; structural stability again guarantees that as long as the gaussian is narrow the dynamics remain qualitatively unchanged) the time evolution of the set of Fock-space amplitudes $\{ c_{z}(t) \}$ is found by solving Schr\"{o}dinger's equation $\mathrm{i} \hbar \partial_{t} \vert \Psi (t) \rangle = \hat{H} \vert \Psi (t) \rangle $ numerically with the Hamiltonian given in Eq.\ (\ref{eq:Hamiltonian-with-angular-momentum}). 

A slice through the resulting quantum probability distribution at time $1.5 \pi /\omega_{\mathrm{pl}}$ is compared against the classical result (thick green line) in Fig.\ \ref{fig:time slice comparing classical quantized decoherence3}(a)  (see also Fig.\ 1 in \cite{Odell2012}). The shape of the discrete probability density $\vert c_z \vert^2$ has recognizable features in common with the classical distribution, including peaks where the slice crosses the caustics (fold lines), but also displays \textit{interference} causing it to oscillate around the classical value. Crucially, the quantum distribution is always finite whereas the classical distribution diverges as the inverse square root of the distance from 
the caustic (as expected \cite{Berry1981}). In fact, in the quantum theory the fold caustics are decorated by Airy functions although these are hard to see here because the Airy functions from different caustics interfere. As will be discussed in more detail in Section \ref{sec:arnold}, in the semiclassical regime $N \gg 1$ the main peaks of the Airy function come to dominate the probability distribution slice, growing as  \cite{Odell2012} 
\begin{equation}
\label{eq:arnoldfold}
\vert \Psi \vert^2_{\mathrm{max}} \propto N^{1/3}
\end{equation}
However, the ``brightest'' parts of the entire pattern are the regions around the cusp points which are decorated by functions known as Pearcey functions \cite{NIST},  and grow as \cite{Mumford2019}
\begin{equation}
\label{eq:arnoldcusp}
\vert \Psi \vert^2_{\mathrm{max}} \propto N^{1/2} \ .
\end{equation}
The exponents $1/3$ and $1/2$ determine how the caustic diverges as the classical limit $N \rightarrow \infty$ is approached \cite{Berry1981}. 

Another feature of the semiclassical regime is that each of the double-peaks in Fig.\ \ref{fig:time slice comparing classical quantized decoherence3} becomes narrow. As $N$ gets large the two fold lines therefore correspond to approximate Schr\"{o}dinger cat states, i.e.\ superpositions of two macroscopically different number states on the two sides of the Josephson junction \cite{Krahn2009}. Related twin Fock states have been made in experiments on spatially separated atomic clouds \cite{Fadel2018,Kunkel2018,Lange2018}.

\section{Master equation}
\label{sec:master}

The effect of decoherence due to a weak continuous measurement of $z$ can be incorporated into the quantum theory via the master equation \cite{Corney1998,Ruostekoski1998}
\begin{equation}
\frac{\partial \hat{\rho}}{\partial \tau} 
=i\left[\hat{J}_x,\hat{\rho} \right]-i\frac{\Lambda}{N}\left[\hat{J}_z^2,\hat{\rho}\right]-D\frac{\Lambda}{N}\left[\hat{J}_z,\left[\hat{J}_z,\hat{\rho}\right]\right],
\label{eq:master-eqn-non-dimensionalized}
\end{equation}
where $D$ governs the strength of the measurement. $\hat{\rho}(\tau)$ is the density operator\textcolor{black}{,} which can be expanded in the Fock ($\hat{J}_{z}$) basis as $\hat{\rho}(t)=\vert \Psi(t)\rangle \langle\Psi(t)\vert=\sum_{q,z} \rho_{q,z}(t)\vert q \rangle \langle z \vert$. The diagonal elements give the populations $\rho_{z,z}\left(t\right)$ of the Fock states as plotted in Fig.\ 
\ref{fig:Classical-trajectories-MF-normal-and-tilt-v2}(c) and Fig.\ \ref{fig:time slice comparing classical quantized decoherence3} (b) and (c). The populations $\rho_{z,z}\left(t\right)$ are the quantum equivalent of the classical probability distribution for the $z$ variable as given by the projection of the classical phase space density onto the $z$-axis: $\int_{-\pi}^{\pi} \rho_{\mathrm{cl}}(\phi,z) \, \mathrm{d} \phi$. When $D=0$, Eq.\ (\ref{eq:master-eqn-non-dimensionalized}) reduces to the quantum Liouville equation $\partial \hat{\rho}/\partial t=-i[\hat{\rho},\hat{H}]$, which is in turn the quantum version of the classical Liouville equation $\partial \rho_{\mathrm{cl}}/\partial t=- \{ \rho_{\mathrm{cl}} , H \}$.

The master equation in Eq.\ (\ref{eq:master-eqn-non-dimensionalized}) is in Kossakowski-Lindblad form, which ensures that the density matrix is positive-definite at all times (see Appendix \ref{app:Lindblad form}) \cite{Lindblad1976,Gorini1978}. The double commutator is responsible for measurement-induced decoherence, which suppresses the density matrix's off-diagonal elements in the number-difference basis due to the gain in information about the number difference by the measurement.  Experimentally, the atom number can be counted continuously (non-destructively) in time using phase contrast imaging \cite{Bradley1997,Andrews1996,Andrews1997}. Other techniques have also been suggested, such as homodyne detection when one of the two BECs is placed inside an optical cavity  \cite{Corney1998}. As long as the measurements are not projective, off-diagonal long range order between the two wells can be preserved \cite{Ruostekoski1998}.

It is important to realize that the populations $\rho_{z,z}\left(t\right)$ given by solving the master equation are probabilities, i.e.\ correspond to the average over many experimental runs.  The measurement records of individual experimental runs are inherently stochastic but  can be simulated by ``unravelling'' the master equation into an ensemble of quantum trajectories obtained by solving a stochastic differential equation or master equation  \cite{Gardiner85,Dalibard92,GisinPercival1992,Carmichael93}; averaging over many quantum trajectories or discarding the measured values then reproduces the predictions of Eq.\ (\ref{eq:master-eqn-non-dimensionalized}). The singularities we discuss in this paper therefore do not appear in any single experimental run but only in the probability distribution. For a relevant example of the quantum trajectory method we point the reader to the paper by Naghiloo \text{et al} \cite{Naghiloo17}.

\begin{figure*}[ht]
	\includegraphics[width=\textwidth]{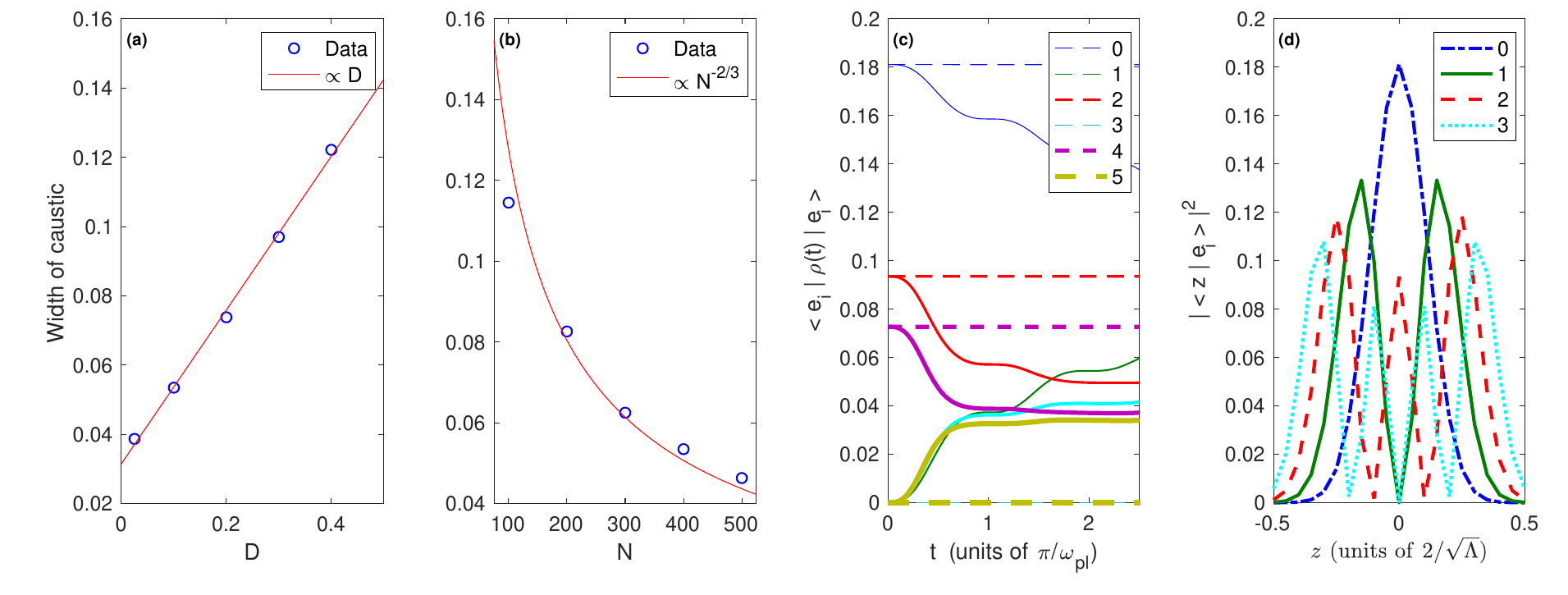}
				\caption{\textbf{(a-b)} Effect of varying the decoherence strength $D$ and the number of particles $N$ on the width of a caustic in the quantum probability distribution. The caustic in question is located at $t=1.5 \pi/ \omega_\mathrm{pl}$ near $z=-0.5$, where $\Lambda=25$. We measure the full width at three-quarters of the maximum of the peak because the half-maximum is too low for large values of $D$ and small values of $N$. \textbf{(a)} For $N=400$, the width of the peak increases linearly with $D$. \textbf{(b)} For $D=0.1$, the width of the peak decreases as the power law $N^{-2/3}$. This is in line with the expected scaling of the width of a fold catastrophe \cite{Berry1981}. All of these trends are maintained if we vary the value of $t$ but track the same fold line.  \textbf{(c-d)} The effect of decoherence on eigenstate populations, for $N=100$ and $\Lambda=25$. \textbf{(c)} Overlaps between $\hat{\rho}$ and the eigenstates of the Hamiltonian [Eq.\ (\ref{eq:Hamiltonian-with-angular-momentum})] for the first six eigenstates (labeled 0--5 and with increasing line thickness). Dashed lines: coherent evolution ($D=0$). Solid lines: with decoherence ($D=0.1$). The coherent evolution couples only to even eigenstates, so the dashed lines representing the populations of states 1, 3, and 5 remain zero for all time whereas decoherence allows the odd eigenstates' populations to grow with time. \textbf{(d)} Probability distributions of the first four eigenstates of the Hamiltonian. The ground state is even, and the states alternate in parity.}
		\label{fig:widths of caustics}
	\end{figure*}

\section{Dynamics with decoherence}
\label{sec:decoherence}

Decoherence is known to smooth rapidly oscillating terms in Sch\"{o}dinger dynamics \cite{Kosloff2013}. In Fig.\ \ref{fig:time slice comparing classical quantized decoherence3} we see that as $D$ is increased the oscillations are progressively damped out and in this sense the system becomes more classical. However, we also see that the caustic peaks are softened too, signalling a departure from the naive classical result given by the TWA. By contrast, increasing $N$ increases the sharpness of the caustics, increasing the similarity to the classical distribution. These observations are quantified in Fig.\ \ref{fig:widths of caustics},  where the width of the caustic is found to increase linearly with $D$ and decrease as a fractional power law with $N$. Furthermore, 
if the root mean square of the difference between the quantized probability distribution and the classical one is computed, one finds that this also decreases as a power law in $N$ (see Appendix \ref{app:quantifying classicality}). This suggests that the required limit for classicality, if defined as the TWA distribution, is $D \rightarrow 0$ and $N \rightarrow \infty$ but such that $D \times N^{\gamma}$ is always finite,  where $\gamma$ is some exponent.

Another interesting limit is the long-time limit for fixed $D$ and $N$. In this case the density matrix becomes diagonal and the probability distribution tends to a steady state that is flat (see Appendix \ref{app:D1-steady-derivation})
\begin{equation}
\left\langle q\left|\hat{\rho}_\mathrm{steady}\right|z\right\rangle\equiv{\rho}_{\mathrm{steady}\,q,z}=\frac{\delta_{q,z}}{N+1} \ ,
\label{eq:steady-state-eqn}
\end{equation}
corresponding to an equal probability of occupying any Fock state.
This behavior can be seen in Fig.\  \ref{fig:Classical-trajectories-MF-normal-and-tilt-v2}(c) where the caustics gradually blur out and dissipate over time.

In order to provide an intuitive explanation of the above results, let us examine what the decoherence term (double commutator) in the master equation is actually doing.  (In formulating the following arguments we have been influenced  by a study on the continuous measurement of the position of a particle in a double well potential presented in \cite{Gagen93}.)  It should first be noted that as long as the tilt term $\Delta E$ is zero, the eigenstates of the Hamiltonian in Eq.\ (\ref{eq:Hamiltonian-with-angular-momentum}) have a well defined parity in Fock space, being either even or odd (the ground state is even, first excited state is odd, second excited state is even and so on alternating up the spectrum). The initial quench excites only the \textit{even} eigenstates. In the absence of decoherence the system will evolve as a coherent superposition of these eigenstates and thus have even symmetry about the center point $z=0$. However, the decoherence term in Eq.\ (\ref{eq:master-eqn-non-dimensionalized}) breaks this symmetry and couples in the odd eigenstates. A linear combination of an even and an odd eigenstate produces a wave function biased to one or the other side of $z=0$. Thus, the  the measurement backaction can be thought of as effectively adding a stochastically varying tilt term to the time evolution. Indeed, as information continuously flows out of the system about the number difference $z$, any non-zero value of $z$ will tilt the probability distribution to one side or the other mimicking the effect of an actual tilt $\Delta E$.

How does the measurement lead to a smoothing out of oscillations in the probability distribution? The double-commutator causes a decay in the probability of finding the system in an even energy eigenstate, and increases the probability of finding it in an odd energy eigenstate, as shown in Fig.\ \ref{fig:widths of caustics}(c). Now, a single eigenstate leads to a probability distribution that oscillates in $z$ with nodes between peaks: in order to obtain a smooth non-oscillating distribution we require many eigenstates be populated, and, in particular, that adjacent eigenstates be populated. This is because adjacent eigenstates in energy have adjacent peaks in number-difference space as shown in Fig.\ \ref{fig:widths of caustics}(d), and thus it is necessary to combine both even and odd eigenstates to fill in the gaps. In this way decoherence smooths the oscillations and broadens the peaks in comparison to the coherent system's probability distribution.

Another way to understand the effect of the double commutator term is to map the master equation onto a Fokker-Planck style equation, which is the equation of motion for the probability distribution.  Choosing the Glauber-Surdashan quasiprobability distribution in phase space $P(\phi,z)$ (the Glauber-Surdashan quasiprobability is positive if the quantum system has a classical analogue), we obtain  
\begin{equation}
\begin{aligned}
\frac{\partial P}{\partial \tau}&=
4\left[-\left(\frac{\partial h}{\partial \phi}\right)\partial_z+\left(\frac{\partial h}{\partial z}\right)\partial_\phi+2\frac{D\Lambda}{N}\partial_\phi^2\right]P ,
\label{eq:diffusioneq}
\end{aligned}
\end{equation}
where $h=\Lambda z^2/2-\sqrt{1-z^2}\cos\phi$ is the dimensionless Hamiltonian, see Appendix \ref{app:FokkerPlanck} for the derivation.   The authors of Reference \cite{Javanainen2013} obtained a very similar equation for the Wigner distribution and pointed out that the last term, which takes the form of a diffusion term in $\phi$, arises as the backaction due to a measurement  of the conjugate variable $z$.  Diffusion can of course be expected to eliminate quantum interference such that the  probability distribution settles to the classical distribution, and also broadens the peaks at the caustics such that the resultant probability distributions are non-singular. It is also significant that the diffusion constant $D \Lambda /N$ depends linearly on $D$ and inversely upon $N$, supporting our earlier observations that these two parameters have opposing effects upon the width of caustics. 

The diffusion equation given in Eq.\ (\ref{eq:diffusioneq}) for the quasiprobability distribution under the influence of decoherence implies another method for smoothing the caustics: the ad hoc addition of diffusive noise terms to the classical Liouville equations. However, classical noise can be reduced by technical improvements, e.g. by going to very low temperatures as in the experiments with ultracold atoms. By contrast, quantum mechanics introduces fundamental noise that cannot be removed so we shall not pursue these alternative approaches here.

\section{Modification of the Arnol'd exponent by decoherence}
\label{sec:arnold}

In the wave theory of caustics divergent singularities are replaced by interference patterns which smooth catastrophes at the scale of the wavelength. Each class of catastrophe is associated with its own characteristic wave function, and for the catastrophes of relevance to us (folds and cusps), these can be expressed as  \cite{Berry1981}
\begin{equation}
	\Psi(\mathbf{C})  \propto k^{1/2} \int_{-\infty}^{\infty}   \mathrm{d} s\;\mathrm{e}^{\mathrm{i} k \, \Phi(s;\mathbf{C}) } 
	\label{eq:generalwavecatastrophe}
 \end{equation}
where $k$ is the wavenumber and $\mathbf{C}=\{ C_{1},C_{2},\ldots \}$ are control parameters representing coordinates and other parameters. $\Phi(s;\mathbf{C})$ is the generating function and for folds it takes the form of a cubic polynomial $\Phi_{1}(s;C)=s^3/3+Cs$, where $C=C(z,t)$. This gives a wave catastrophe of the form 
\begin{equation}
\label{eq:airywc}
\Psi_{1}(C) \propto (2 \pi k^{1/6}) \  \mathrm{Ai}(k^{2/3} C)
\end{equation} 
where $\mathrm{Ai}(x)$ is the Airy function \cite{NIST}
\begin{equation}
\mathrm{Ai}(x)=\frac{1}{2 \pi} \int_{-\infty}^{\infty} \exp [\mathrm{i}(s^3/3+xs)] \mathrm{d}s \ .
\end{equation}
Cusps have a quartic generating function featuring two control parameters $\Phi_{2}(s;C_{1},C_{2})=s^4/4+C_{2} s^2/2 +C_{1} s$,  giving
\begin{equation}
\label{eq:pearceywc}
\Psi_{2}(C_{1},C_{2}) \propto (2 \pi k^{1/4})  \ \mathrm{Pe}(C_{1}k^{3/4},C_{2}k^{1/2}) 
\end{equation}
where $\mathrm{Pe}(x,y)$ a complex function of two variables known as the Pearcey function \cite{NIST} 
\begin{equation}
\mathrm{Pe}(x,y)=\frac{1}{2 \pi} \int_{-\infty}^{\infty} \exp [\mathrm{i}(s^4/4+y s^2/2+x s)] \mathrm{d}s.
 \end{equation}
 The higher catastrophes contain the lower ones: away from its tip the cusp evolves into two fold lines. The same is true of wave catastrophes: away from the cusp tip at $x=y=0$, the Pearcey function 
 evolves into two Airy functions whose tails overlap in the middle.
 
 It is evident from the above expressions that wave catastrophes have remarkable scaling properties as $k$ is varied. The powers of $k$ multiplying the control parameters (coordinates) in the arguments of the Airy and Pearcey functions show that varying $k$ is equivalent to varying the fringe spacing of the interference patterns. For example, if $k$ is increased (smaller wavelength) the fringe spacing of the interference pattern dressing a fold catastrophe decreases as  $k^{-2/3}$. In our case the role of $k$ is played by the total number of atoms $N$ \cite{Mumford2019} and we have in fact already seen in Fig.\ \ref{fig:widths of caustics}(b) that the width of the main peak  decreases as $N^{-2/3}$, in agreement with the wave catastrophe prediction. This is, therefore, a purely coherent wave effect.

\begin{figure}[t]
	\includegraphics[width=8cm]{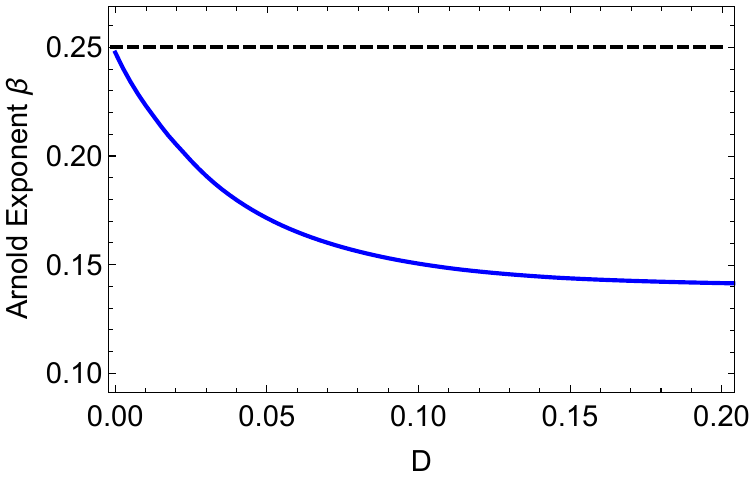}
		\caption{Blue curve: dependence of the Arnol'd scaling index $\beta$ (evaluated at the cusp points) on the strength of the decoherence $D$. This curve is evaluated for $\Lambda=8$, but we find the same trend for all values of $\Lambda$. At $D=0$ we recover the standard wave catastrophe prediction for a cusp of $\beta=1/4$.}
		\label{fig:openarnold}
	\end{figure}

For present purposes, the most interesting scaling is that of the amplitude as this seems to be more strongly affected by decoherence. We have previously given, in Eqns.\ (\ref{eq:arnoldfold}) and (\ref{eq:arnoldcusp}), the wave catastrophe predictions for the scaling of the peaks $\vert \Psi \vert^2_{\mathrm{max}}$ of the probability distribution with $N$. Now we can see where these results come from: they are (twice) the amplitude scaling given in Eqns.\ (\ref{eq:airywc}) and  (\ref{eq:pearceywc}), respectively. The exponents $\beta=1/6$ and $\beta=1/4$ are known as the Arnol'd exponents and show how the wave catastrophe diverges $\Psi_{\mathrm{max}} \sim N^{\beta}$ in the classical limit $N \rightarrow \infty$ \cite{Berry1981}.  In Fig.\  \ref{fig:openarnold} we plot the dependence of $\beta$ on measurement strength $D$. The curve was obtained by finding the highest peak in the time-dependent probability distribution $\vert \Psi(z,t) \vert^2$ from the numerical solution of the master equation (excluding very early times where the dominant peak comes from the initial condition of starting in a single Fock state) and tracking the height of this peak as $N$ is varied.  We find that at $D=0$ we recover the result for a cusp, namely $\beta=1/4$, but the effect of decoherence is to reduce $\beta$ making the catastrophe less singular even though it is at the same time making the system more classical. Once again, this result illustrates the dual effects of decoherence in removing quantum interference but also adding noise that softens classical singularities.

\section{Conclusions} 
\label{sec:conclusions}

Classical theories breakdown at caustics where they predict singularities. Furthermore, catastrophe theory predicts that caustics occur generically, without the need for fine tuning or symmetry, due to their structural stability. In many-body theories caustics can occur in more subtle ways, such as the probability distribution in Fock space, rather than the more tangible coordinate space, but nevertheless this represents a striking failure of classical field theory. If decoherence merely reduced the quantum prediction to the classical one this would spell trouble because true singularities would exist even in finite systems ($N \neq \infty$). Caustics therefore represent a particularly stern test of decoherence theory. 

In this paper we compared and contrasted numerical solutions of the master equation describing a continuously measured  many-particle two-mode quantum system against those of the classical (GPE) and quantum (Bose-Hubbard) theories for the equivalent closed system. We find that continuous weak measurements do not lead us precisely to the closed classical result and instead measurement backaction introduces noise that softens the singularities, thereby resolving any potential paradox. 

The results one obtains in such quantum-to-classical transitions will depend sensitively on the order in which the limits $N \rightarrow \infty$ and $D \rightarrow 0$ are taken.  At any finite $N$ the results of decoherence of a quantum system must be finite. However, in the thermodynamic limit ($N \rightarrow \infty$) singularities are allowed, although without decoherence the quantum probability distribution will oscillate infinitely fast. In the `old' way of taking the classical limit these oscillations were presumed averaged out by the finite resolution of a detector \cite{Berry1972}, but with decoherence a range of classical limits exist by letting $N \rightarrow \infty$ while keeping $D>0$. The degree to which caustics are softened will depend on the magnitude of $D$. One way to quantify this is through the Arnol'd scaling exponents: we find that the Arnol'd exponent for the amplitude of cusp points is reduced as $D$ increases. 

Since caustics are formed during dynamics,  the Arnol'd exponent can be considered as a type of critical exponent for non-equilibrium quantum dynamics analogous to the critical exponents that describe equilibrium phase transitions. Both situations fundamentally involve singularities, and in both cases the exponents depend on the class of singularity. It is noteworthy in this context that critical exponents for phase transitions in open quantum systems can be modified from their closed values. This has, for example, been predicted  \cite{Nagy2011,Bhaseen2012,Mumford2015} and observed \cite{Brennecke2013} for the Dicke phase transition. We propose that the modification of the Arnol'd exponent in an open quantum system is a dynamical analogue of this type of effect.

Finally, we would like to point out that although we have focused here on the scalar bosonic Josephson junction/two mode Bose-Hubbard model,  our Hamiltonian in its spin form, Eq.\ (\ref{eq:Hamiltonian-with-angular-momentum}), is exactly that of the TFIM with infinite-range interactions \cite{Das06}. Thus, the physics described here can also be realized in spin systems with long-range interactions such as trapped ions \cite{Britton12,Jurcevic14,Richerme14,Bohnet16,Safavi18} or Rydberg atoms \cite{Ashida19} where two internal states form the spin degrees of freedom.  Controllable decoherence can be achieved in these systems through continuous measurement of the difference between the number of spin-up and spin-down spins (which can be read out using a global fluorescence measurement \cite{Bohnet16}), in an exactly analogous fashion to the scheme discussed in Section \ref{sec:master}. Alternatively, in the case of crystals of ions in a Penning trap, the spins are coupled to a bath of phonons which can act as the decohering environment \cite{Safavi18}. These phonons, which play the role of photons in an analogue of the Dicke model, mediate the long-range interactions between spins and are activated by pairs of lasers allowing for a high degree of control. 

Concerning future directions, we note that both the infinite-range TFIM and the Dicke model undergo $\mathbb{Z}_2$ symmetry breaking quantum phase transitions between paramagnetic/normal states and ferromagnetic/superradiant states \cite{Das06,Garraway11,buonsante12}. However, there is a difference:  the Dicke model is chaotic (i.e.\ in the classical limit the dynamics are chaotic) in the superradiant phase whereas the infinite-range TFIM is integrable and hence has regular dynamics in both phases \cite{lambert04,mumford14a,mumford14b}.
An interesting question concerns the fate of caustics in the chaotic phase: structural stability guarantees they must survive if the chaos is weak (this is essentially the famous Kolmogorov-Arnol'd-Moser theorem which says that most of the tori in phase space, which characterize integrable systems, survive in the presence of small nonintegrable perturbations \cite{Arnold89}), but strong chaos must eventually disrupt the whorl structures in phase space (Fig.\ \ref{fig:phasespacedynampic} in Appendix \ref{app:phasespace}) so thoroughly that the caustics should melt away, possibly evolving into scars \cite{Choi19}.

\acknowledgements
We thank Professor J. Ruostekoski for discussions and an anonymous referee for useful suggestions. We also acknowledge funding from the Natural Sciences and Engineering Research Council of Canada. This work is supported in part by the M. Hildred Blewett Fellowship of the American Physical Society, www.aps.org.

\begin{appendix}
	
	\vspace{3ex}

	\section{Dynamics in classical phase space}	
\label{app:phasespace}

The Josephson equations of motion given in Eq.\ (\ref{eq:Hamilton's equations}) correspond to those of a classical pendulum of variable length, i.e.\ a length which depends on the angular momentum represented by $z$ through the square root factor $\sqrt{1-z^2}$  \cite{Smerzi1997}. Apart from the oscillatory and rotational motion that is present in the standard rigid pendulum, such a non-rigid pendulum can also sustain a periodic motion about its inverted position known as $\pi$-oscillations \cite{Raghavan1999}. However, when $\Lambda>1$ as here, the $\pi$-oscillations disappear \cite{Mumford2017}. Furthermore, the initial conditions we use in this paper, namely $z=0$ and $\phi$ drawn uniformly from the range $\{0 \ldots 2 \pi \}$, do not excite the rotational modes. We are therefore exclusively exciting the standard oscillatory modes (known as plasma oscillations in the Josephson junction literature) which appear at low energies as ellipses in phase space.

\begin{figure}
		\includegraphics[width=\columnwidth]{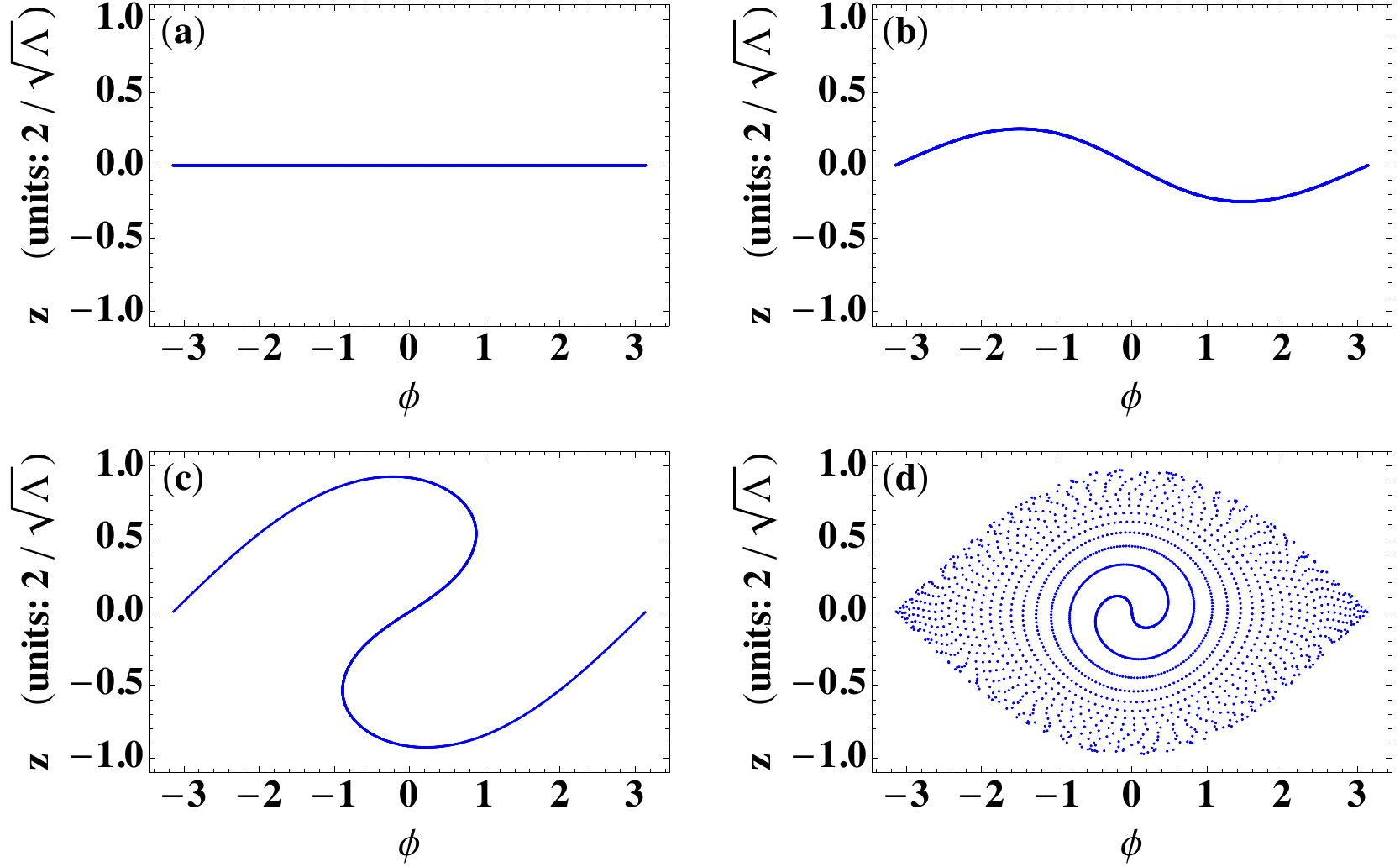}
		\caption{Dynamics in classical phase space $(\phi,z)$ obtained by numerically solving Josephson's equations as given in Eq.\ (\ref{eq:Hamilton's equations}) with $\Lambda=25$ and $\Delta E=0$. Each panel is for a different elapsed time: $\tau=\{0, 0.1,0.5,20\}$ and is made up of 1001 points corresponding to different initial conditions.  As explained in Appendix \ref{app:phasespace}, the dynamics cause the initial distribution to wind up into a whorl, the stationary points of which give rise to the successive caustics seen in Fig. \ref{fig:Classical-trajectories-MF-normal-and-tilt-v2}(a). Projecting onto the $z$ axis gives the classical probability distribution given by the green solid line in Fig.\ \ref{fig:time slice comparing classical quantized decoherence3}(a). At long times the phase space distribution is spread ergodically (but non-uniformly) over phase space as discussed in Reference \cite{BerryODell1999}.}
		\label{fig:phasespacedynampic}
	\end{figure}

The larger the initial value of $\phi$ (i.e.\ the larger the initial value of the energy), the greater the period of the pendulum. This basic feature of the motion of a pendulum generates a flow in phase space that leads to whorls as shown in Fig.\ \ref{fig:phasespacedynampic}. Each panel in this figure is made up of 1001 points, each point being a pendulum with a different initial value of $\phi$. In panels (a), (b) and (c) in Fig.\ \ref{fig:phasespacedynampic} the points are so dense they form what appears to be a solid curve and only at longer times do the individual points become evident. In particular,   Fig.\ \ref{fig:phasespacedynampic}(a) shows the ensemble of initial conditions at $\tau=0$ which give a straight line along $z=0$,  and Fig.\ \ref{fig:phasespacedynampic}(b) shows a short time later, $\tau=0.1$, where the line has evolved into a curve. By the time $\tau=0.5$, shown in Fig.\ \ref{fig:phasespacedynampic}(c), the higher frequency motion of the low energy pendula near the origin of phase space relative to the lower frequency motion of the higher energy pendula further out causes the curve to evolve into a whorl \cite{BerryODell1999}. The stationary points of this whorl as a function of $\phi$ give rise  to caustics in the $z$ variable and vice versa, as can be seen by projecting the phase space distribution on to either the $z$ or $\phi$ axis, respectively. As time evolves further, the curve continues to wind up, generating more and more caustics. By the time $\tau=20$, shown in Fig.\ \ref{fig:phasespacedynampic}(d), the different points are spread out over the energetically allowed region of phase space which is bounded by the separatrix $E=E_{J}$. In fact, the motion is ergodic because in one-dimensional systems such as this the dynamics at each value of the energy explores the entire available energy surface (which corresponds to an ellipse at low energy but is shaped like an eye near the separatrix). Using this insight, an expression for the phase space density at long times can be derived as shown in Reference \cite{BerryODell1999}.

	\section{Structural stability of catastrophes: stability to changes in Hamiltonian parameters and initial conditions}
	\label{app:stability to initial conditions}
\begin{figure}
		\includegraphics[width=\columnwidth]{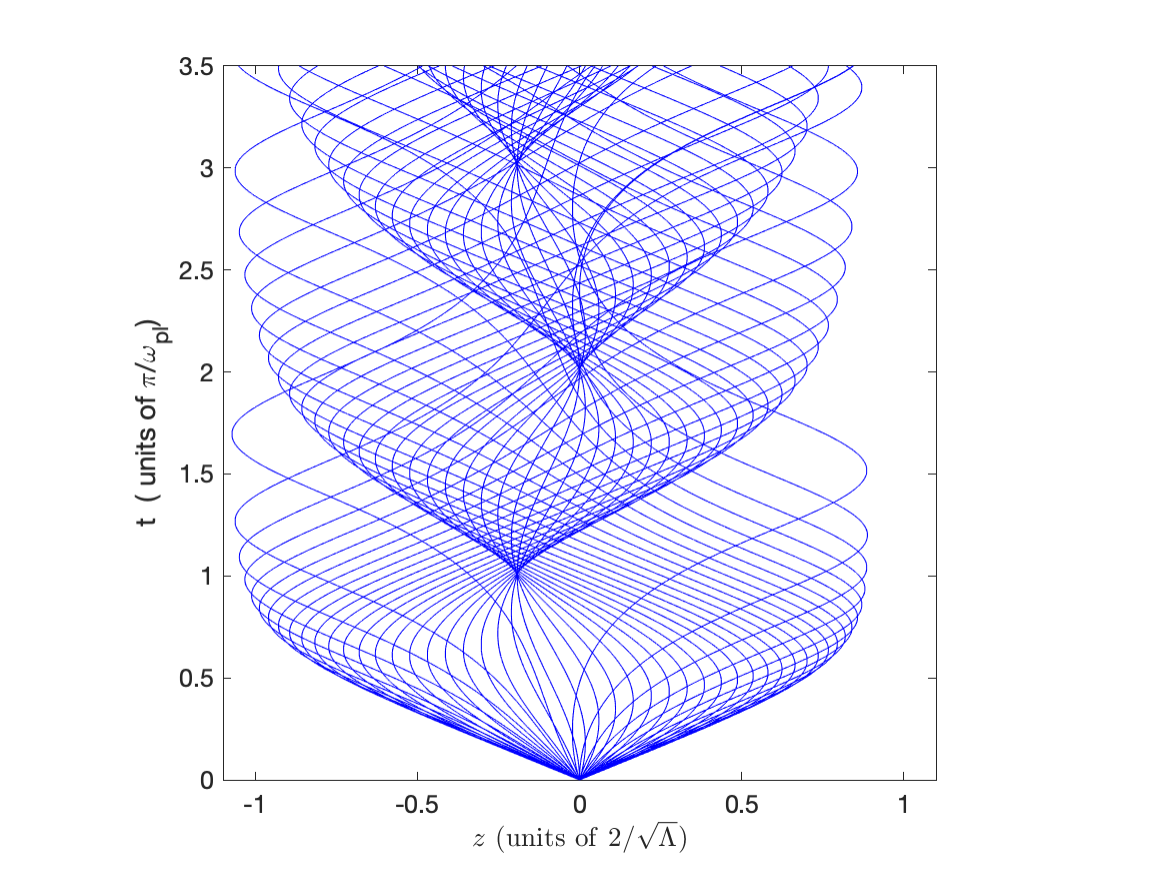}
		\caption{Classical trajectories with the same parameters as Fig. \ref{fig:Classical-trajectories-MF-normal-and-tilt-v2}, except with $\Delta E=1$. Having a finite tilt bias knocks the cusps off axis, but due to their structural stability in two dimensions they remain as cusps.}
		\label{fig:initial-tilt}
	\end{figure}

	We evolve the Josephson equations using a tilt term with the fiducial value $\Delta E=1$. Using the same parameters as in Fig.\ \ref{fig:Classical-trajectories-MF-normal-and-tilt-v2}, the tilt bias knocks the cusps off of the $z=0$ axis as can be seen in Fig.\ \ref{fig:initial-tilt}. However, caustics still form and retain their qualitative properties, but with their cusps oscillating between $z=0$ and $z=-0.2$.

\begin{figure}\includegraphics[width=\columnwidth]{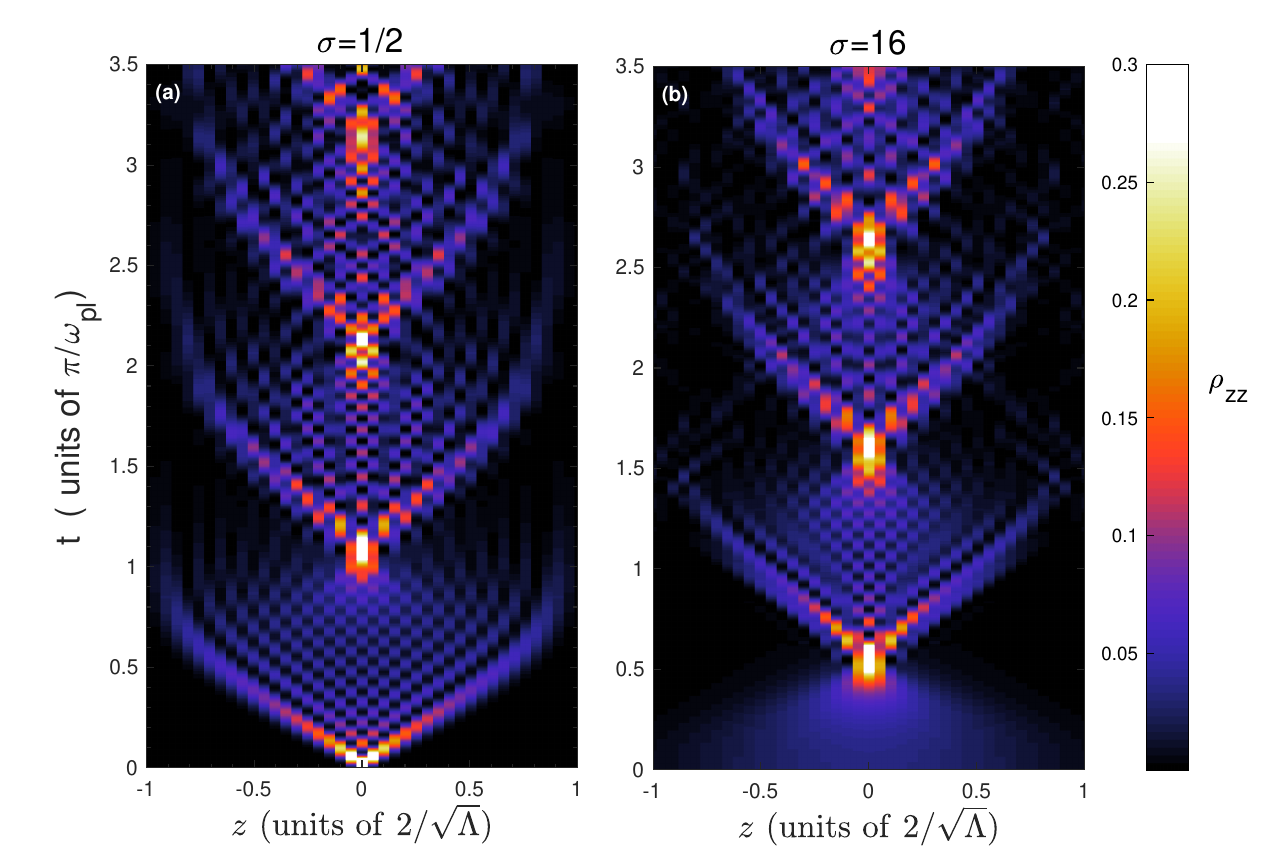}
		\caption{Stability of a quantum catastrophe to varying the width of the initial state in Fock space. Evolution of the $100$-particle system with $\Lambda=25$, in the initial state given by Eq. (\ref{eq:gaussian-initial-state-eqn}). \textbf{(a)} With the initial width $\sigma=1/2$, the \textcolor{black}{behavior} is quite similar to the $z=0$ initial condition case. \textbf{(b)} Increasing the initial width to $\sigma=16$ again yields the catastrophe structures, which now begin at $t=0.5\pi/\omega_\mathrm{pl}$.}
		\label{fig:initial-gaussians}
	\end{figure}

	Next, we examine the effect of taking the initial state in the second-quantized formulation to be a Gaussian superposition of states  rather than a single Fock state:
	\begin{equation}
	c_z=\frac{1}{Z}e^{-\frac{z^2 N^2}{8\sigma^2}},
	\label{eq:gaussian-initial-state-eqn}
	\end{equation}
	where
	\begin{equation}
	Z=\sqrt{\sum_{z}c_z^2}.
	\end{equation}
	These initial conditions are then evolved using the Schr\"{o}dinger equation with the Hamiltonian Eq. (\ref{eq:Hamiltonian-with-angular-momentum}) exactly as in the case of the single Fock state.
	For values of $\sigma$ up to $1/2$, the behavior is nearly identical to starting in the $z=0$ state, see Fig.\ \ref{fig:initial-gaussians}(a). Since these initial conditions are qualtitatively similar to $\left|0\right\rangle$, this further verifies the stability prediction of catastrophe theory. 
	
	Large $\sigma$ values correspond to starting in a state with maximal uncertainty in $z$ but minimal uncertainty in $\phi$, and the initial cusp moves to a later time $t>0$, as seen in Fig.\ \ref{fig:initial-gaussians}(b), but still occurs. We note that the pixelation in Fig.\ \ref{fig:initial-gaussians} is not a resolution limit, but rather the result of $z$ being quantized. It is this very quantization that corrects the classical results through a quantized Airy function.
	
	\section{Form of the master equation}
	\label{app:Lindblad form}
	We here show that Eq.\ (\ref{eq:master-eqn-non-dimensionalized}) is in Kossakowski-Lindblad form. This is the most general form that ensures the positivity of the reduced density matrix $\hat{\rho}_S$ at all times, where
	\begin{equation}
	\hat{\rho}_S=\mathrm{Tr}_E\left[\hat{\rho}_{SE}\left(t\right)\right],
	\end{equation}
	and $S$, $E$, and $SE$ denote the system, environment, and composite system, respectively \cite{Schlosshauer2005}.
	 We rewrite the master equation in this general form by defining the Lindblad superoperator $\mathcal{L}$
	\begin{equation}
	\mathcal{L}\hat{\rho}= \hat{L}\hat{\rho }\hat{L}^\dagger-\frac{1}{2}\left(\hat{\rho }\hat{L}^\dagger \hat{L}+ \hat{L}^\dagger \hat{L}\hat{\rho}\right).
	\end{equation}
	With 
	\begin{equation} \hat{L}= \hat{L}^\dagger\equiv\sqrt{2D\Lambda/N} \hat{J}_z\end{equation}
	 and the reduced Hamiltonian 
	 \begin{equation}\tilde{\hat{H}}\equiv-\hat{J}_x+\frac{\Lambda}{N} \hat{J}_z^2,
	 \end{equation}
	we see that the promised equation
	\begin{equation}
	\dot{\hat{\rho}}=-i\left[\tilde{\hat{H}},\hat{\rho}\right]-\mathcal{L}\hat{\rho} 
	\end{equation}
	is equivalent to our master equation
	\begin{equation}
	\frac{\partial \hat{\rho}}{\partial \tau} 
	=i\left[\hat{J}_x,\hat{\rho}\right]-i\frac{\Lambda}{N}\left[\hat{J}_z^2,\hat{\rho}\right]-D\frac{\Lambda}{N}\left[\hat{J}_z,\left[\hat{J}_z,\hat{\rho}\right]\right].
	\end{equation}

		\section{Quantifying the return to classicality}
		\label{app:quantifying classicality}
\begin{figure}[t]
		\includegraphics[width=\columnwidth]{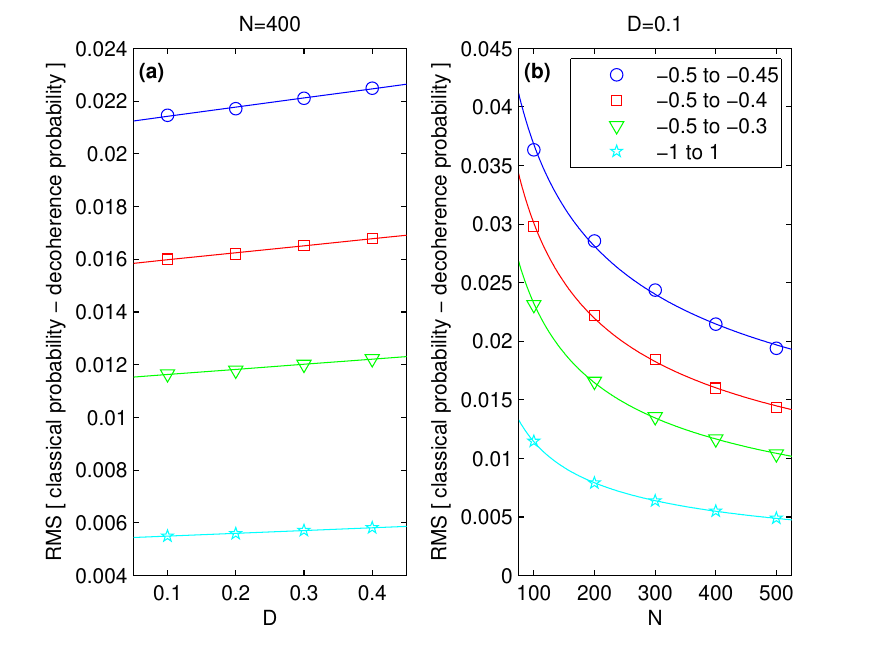}
		\caption{Root-mean-square (RMS) of the difference between the classical probability distribution [Fig.\ \ref{fig:time slice comparing classical quantized decoherence3}(a)] and the quantum distribution for varying decoherence strength and number of particles for $\Lambda=25$ and time $t\omega_\mathrm{pl}=1.5\pi$. The RMS is calculated for various ranges of 
$z$, all of which show the same trends. Since the inside of a caustic is well-studied by catastrophe theory \cite{Berry1981}, we use ranges between the caustic at $z=-0.5$ and the three locations 
			$z=-0.45$ (blue circles), 
			$z=-0.4$ (red squares), and 
			$z=-0.3$ (green triangles), as well as the range 
			$-1\leq z\leq 1$ (cyan stars).
			\textbf{(a)} Here $N=400$. For all ranges of 
			$z$, the difference between the classical probability and the probability with decoherence increases linearly with $D$. The fits are as follows: $-0.5$ to $-0.45$ has $\mathrm{RMS}=0.0035 \times D+0.021$; $-0.5$ to $-0.4$ has $\mathrm{RMS}=0.0027 \times D+0.016$; $-0.5$ to $-0.3$ has $\mathrm{RMS}=0.0019 \times D+0.011$; and, $-1$ to $1$ has $\mathrm{RMS}=0.0011 \times D+0.0054$.
			\textbf{(b)} Here $D=0.1$. For all ranges of 
			$z$, the difference between the classical probability and the probability with decoherence decreases as a power law in $N$. The fits are as follows: $-0.5$ to $-0.45$ has $\mathrm{RMS}=0.22 \times N^{-0.39}$; 
			$-0.5$ to $-0.4$ has $\mathrm{RMS}=0.24 \times N^{-0.45}$; $-0.5$ to $-0.3$ has $\mathrm{RMS}=0.23 \times N^{-0.50}$; and, $-1$ to $1$ has $\mathrm{RMS}=0.13 \times N^{-0.53}$.
			All of the above trends are conserved while varying the value of $t$.}
		\label{fig:rms classical minus decoherence}
	\end{figure}

		Here we investigate further the effects of varying decoherence strength $D$ and particle number $N$ on obtaining the classical limit from the system with decoherence.  We use the root-mean-squared (RMS) difference between the classical and quantum probability distributions to measure the return to classicality, ensuring that our measurement is not obscured by an averaging effect. Fig. \ref{fig:rms classical minus decoherence} plots this quantity versus $D$ and $N$ for various ranges in $z$. Three of the ranges focus on the inside edge of the caustic, and the fourth looks at the entire range $-1\leq z\leq 1$. We find that the RMS value increases linearly with $D$ and decreases as a power-law in $N$, the same trend as the widths of the peaks in Fig. \ref{fig:widths of caustics}. The trend is seen not only on the inside edge of the caustic, but also across the entire range $-1\leq z\leq 1$. We see again that increasing decoherence slowly pushes the system away from the classical distribution, and that increasing the number of particles while decoherence is turned on rapidly pushes the system toward classicality.

	\section{Steady state derivation}
	\label{app:D1-steady-derivation}
	
	The master equation can be expressed in the number-difference basis as
	\begin{equation}
	\begin{aligned}
	\frac{\mathop{d}\rho_{q,z}}{\mathop{d} \tau}= & i\frac{N\Lambda}{4}\rho_{q,z}\left(z^2-q^2\right)\\
	& -i\frac{N}{4}\left[\rho_{q,z+1}\sqrt{\left(1+z+\frac{2}{N}\right)\left(1-z\right)} \right. \\ 
	& \left.+ \rho_{q,z-1}\sqrt{\left(1-z+\frac{2}{N}\right)\left(1+z\right)} \right. \\
	&\left.- \rho_{q+1,z}\sqrt{\left(1+q+\frac{2}{N}\right)\left(1-q\right)} \right. \\
	& \left.- \rho_{q-1,z}\sqrt{\left(1-q+\frac{2}{N}\right)\left(1+q\right)}\right]\\
	&-D\frac{N\Lambda}{4}\rho_{q,z}\left(q-z\right)^2 .
	\end{aligned}
	\label{eq:master-equation-Jz-basis-no-D2}
	\end{equation}
	Choosing, e.g., 
	$z=q-1$ yields 
	\begin{equation}
	\begin{aligned}
	\dot{\rho}_{q,q-1}= & i\frac{N\Lambda}{4}\rho_{q,q-1}\left(-2q+1\right)\\
	&-i\frac{N}{4}\left[\rho_{q,q}\sqrt{\left(1+q\right)\left(1-q+\frac{2}{N}\right)} \right.\\ 
	& \left.+ \rho_{q,q-2}\sqrt{\left(1-q+\frac{4}{N}\right)\left(1+q-\frac{2}{N}\right)} \right.\\
	&\left.- \rho_{q+1,q-1}\sqrt{\left(1+q+\frac{2}{N}\right)\left(1-q\right)} \right.\\
	& \left.- \rho_{q-1,q-1}\sqrt{\left(1-q+\frac{2}{N}\right)\left(1+q\right)}\right]\\
	&-D\frac{N\Lambda}{4}\rho_{q,q-1} .
		\end{aligned}
	\end{equation}
	
	By inspection of (\ref{eq:master-equation-Jz-basis-no-D2}) with this case, the off-diagonal terms should decay to zero exponentially with increasing $\tau$. 
	Setting the time derivative to $0$ and retaining only the diagonal elements of $\rho$ yields
	\begin{equation}
	\rho_{q,q}=\rho_{q-1,q-1},
	\end{equation}
	and so
	\begin{equation}
	\rho_{\mathrm{steady} \,q,z}=\frac{\delta_{q,z}}{N+1},
	\end{equation}
	where we have employed the normalization condition $\mathrm{Tr}\left[\rho\right]=1$.
	
	Alternatively, one can observe that $\langle q \vert [J_z,[J_z,\rho ]] \vert z \rangle=(J_{z\,q,q}-J_{z\,z,z})^2 \rho_{q,z}$ is only identically zero when $\rho_{q,z}=0$ for all $q\neq z$, ensuring that the steady state is diagonal in the Fock basis. Elementary commutator algebra shows that
	\begin{equation}
	\begin{aligned}
	\left\langle q \vert \left[H,\rho\right] \vert z\right\rangle&=\sum_j H_{q,j}\rho_{j,z}-\rho_{q,j}H_{j,z}\\
	&=H_{q,z}\left(\rho_{z,z}-\rho_{q,q}\right)
	\end{aligned}
	\end{equation}
	for $\rho$ being diagonal. This commutator is only identically zero when $\rho_{q,q}=\rho_{z,z}$, which is the steady state derived above. Reference \cite{Gagen93} found the same steady state for the $N=2$ reduced density matrix.

\section{Derivation of the Fokker-Planck equation}
	\label{app:FokkerPlanck}	

Here we derive a Fokker-Planck equation from the master equation Eq.\ (\ref{eq:master-eqn-non-dimensionalized}) using the P-representation following Breuer \& Pettrucione \cite{Breuer_and_Pettrucione}. The authors of Reference \cite{Javanainen2013} did a calculation for a similar system using the related Wigner function method, but for a slightly different measurement scheme. We define the probability density function $P\left(\alpha,\alpha^*,\beta,\beta^*;t\right)$ by
\begin{equation}
\hat{\rho}\left(t\right)=\int d^2\alpha d^2\beta P\left(\alpha,\alpha^*,\beta,\beta^*;t\right) \left|\alpha\right\rangle\left\langle\alpha\right| \left|\beta\right\rangle\left\langle\beta\right|
\end{equation}
for coherent states
\begin{equation}
\left|\alpha\right\rangle=\exp\left(\alpha \hat{a}_L^\dagger-\alpha^*\hat{a}_L\right)\left|0\right\rangle
\end{equation}
and
\begin{equation}
\left|\beta\right\rangle=\exp\left(\beta \hat{a}_R^\dagger-\beta^*\hat{a}_R\right)\left|0\right\rangle.
\end{equation}
These can be substituted into Eq. (\ref{eq:master-eqn-non-dimensionalized}) using the correspondences
\begin{equation}
\begin{aligned}
\hat{a}_L\hat{\rho}\leftrightarrow\alpha P,\,\hat{a}_L^\dagger\hat{\rho}\leftrightarrow\left(\alpha^*-\partial_{\alpha}\right)P \\
\hat{\rho} \hat{a}_L\leftrightarrow\left(\alpha-\partial_{\alpha^*}\right)P,\,\hat{\rho} \hat{a}_L^\dagger\leftrightarrow\alpha^*P \\
\hat{a}_R\hat{\rho}\leftrightarrow\beta P,\,\hat{a}_R^\dagger\hat{\rho}\leftrightarrow\left(\beta^*-\partial_{\beta}\right)P \\
\hat{\rho} \hat{a}_R\leftrightarrow\left(\beta-\partial_{\beta^*}\right)P,\,\hat{\rho} \hat{a}_R^\dagger\leftrightarrow\beta^*P 
\end{aligned}
\end{equation}
to yield
\begin{equation}
\begin{aligned}
\frac{d P}{d \tau}
&=\left(\frac{i}{2}\left(\alpha^*\partial_{\beta^*}-\alpha\partial_\beta+\beta^*\partial_{\alpha^*}-\beta\partial_\alpha\right) \right.\\
&\,\,\,\,\,\left.-i\frac{\Lambda}{4N}\left[-2\left(\alpha\alpha^*-\beta\beta^*\right)\left(  \alpha\partial_\alpha-\alpha^*\partial_{\alpha^*}+\beta^*\partial_{\beta^*}-\beta\partial_\beta   \right) \right.\right.\\
&\,\,\,\,\,\left.\left.
+\alpha\partial_\alpha-\alpha^*\partial_{\alpha^*}+\beta\partial_\beta-\beta^*\partial_{\beta^*}
\alpha^2\partial_\alpha^2-\alpha^{*2}\partial_{\alpha^*}^2
\right.\right.\\
&\,\,\,\,\,\left.\left.
+\beta^2\partial_\beta^2-\beta^{*2}\partial_{\beta^*}^2 +\alpha\partial_\alpha\beta\partial_\beta-\alpha^*\partial_{\alpha^*}\beta^*\partial{\beta^*}\right]\right.\\
&\,\,\,\,\,\left. +\frac{D\Lambda}{4N}\left\{-\frac{1}{2}\left[\left(\alpha^*\partial_{\alpha^*}-\alpha\partial_\alpha\right)^2+\left(\beta^*\partial_{\beta^*}-\beta\partial_\beta\right)^2\right] \right.\right.\\
&\,\,\,\,\,\left.\left.
+\left(\alpha^*\partial_{\alpha^*}-\alpha\partial_\alpha\right)\left(\beta^*\partial_{\beta^*}-\beta\partial_\beta\right)\right\}\right)P.
\end{aligned}
\label{eq:Fokker-Planck-coherent-variables}
\end{equation}

Next we transform to the real variables $z$ and $\phi$ 
\begin{equation}
\alpha\equiv\sqrt{\frac{N}{2}\left(1-z\right)}\exp\left(-\frac{i\phi}{2}\right),\, \beta\equiv\sqrt{\frac{N}{2}\left(1+z\right)}\exp\left(\frac{i\phi}{2}\right),
\end{equation}
as per Reference \cite{Javanainen2013}. We compute
\begin{equation}
\begin{aligned}
\partial_\alpha&=\frac{2}{\alpha}\left[\left(z-1\right)\partial_z+i\partial_\phi\right]\\
\partial_\beta&=\frac{2}{\beta}\left[\left(z+1\right)\partial_z-i\partial_\phi\right],
\end{aligned}
\end{equation}
noting that $\left(\partial_\alpha\right)^*=\partial_{\alpha^*}$, etc. We substitute into Eq. (\ref{eq:Fokker-Planck-coherent-variables}) to yield
\begin{equation}
\begin{aligned}
\frac{\partial P}{\partial \tau}&=
\left[-4\sqrt{1-z^2}\sin\phi\,\partial_z \right.\\
&\,\,\,\,\,\left.
+4\frac{z}{\sqrt{1-z^2}}\cos\phi\,\partial_\phi+4\Lambda z\partial_\phi+8\frac{D\Lambda}{N}\partial_\phi^2\right]P\\
&=4\left[-\left(\frac{\partial h}{\partial \phi}\right)\partial_z+\left(\frac{\partial h}{\partial z}\right)\partial_\phi+2\frac{D\Lambda}{N}\partial_\phi^2\right]P
\end{aligned}
\end{equation}
for 
\begin{equation}
h=\frac{\Lambda z^2}{2}-\sqrt{1-z^2}\cos\phi.
\end{equation}
\end{appendix}


\begin{thebibliography}{99}

\bibitem{Zurek1991}{W. H. Zurek, Phys. Today \textbf{44}, 36 (1991).}
\bibitem{Zurek2003}{W. H. Zurek, Rev. Mod. Phys. \textbf{75}, 715 (2003). }
\bibitem{Schlosshauer2005}{M. Schlosshauer, Rev. Mod. Phys. \textbf{76}, 1267
(2005).}
\bibitem{Joos85}{E. Joos  and H. D. Zeh, Z. Phys. B \textbf{59} 223 (1985).}
\bibitem{Zurek94}{W. H. Zurek and J. P. Paz. Phys. Rev. Lett. \textbf{72}, 2508 (1994).}
\bibitem{Zurek1993b}{W. H. Zurek, S. Habib, and J. P. Paz, Phys. Rev. Lett. \textbf{70}, 1187 (1993).}
\bibitem{Gallis96}{M. R. Gallis, Phys. Rev. A \textbf{53}, 655 (1996).}
\bibitem{Spiller94}{T. P. Spiller and J. F. Ralph, Phys. Lett. A \textbf{194} A235 (1994).}
\bibitem{Habib98}{S. Habib, K. Shizume, and W. H. Zurek, Phys. Rev. Lett. \textbf{80}, 4361 (1998).}
\bibitem{Bhattacharya2000}{T. Bhattacharya, S. Habib,  and K. Jacobs, Phys. Rev. Lett. \textbf{85}, 4852 (2000). }
\bibitem{Karkuszewski2002}{Z. P. Karkuszewski, C. Jarzynski, and W. H. Zurek, Phys. Rev. Lett. \textbf{89}, 170405  (2002).}
\bibitem{Brune1996}{M. Brune, E. Hagley, J. Dreyer, X. Maitre, A. Maali, C. Wunderlich, J. M. Raimond, and S. Haroche, Phys. Rev. Lett. \textbf{77}, 4887 (1996).}
\bibitem{Myatt2000}{C. J. Myatt, B. E. King, Q. A. Turchette, C. A. Sackett, D.
Kielpinski, W. M. Itano, C. Monroe, and D. J. Wineland, Nature (London) \textbf{403}, 269 (2000).}
\bibitem{Turchette2000}{Q. A. Turchette, C. J. Myatt, B. E. King, C. A. Sackett, D. Kielpinski, W. M. Itano, C. Monroe, and D. J. Wineland, Phys. Rev. A \textbf{62}, 053807 (2000).}
\bibitem{Kokorowski2001}{D. A. Kokorowski, A. D. Cronin, T. D. Roberts, and D. E. Pritchard, Phys. Rev. Lett. \textbf{86}, 2191 (2001).}
\bibitem{Hornberger2003}{K. Hornberger, S. Uttenthaler, B. Brezger, L. Hackerm\"{u}ller, M. Arndt, and A. Zeilinger, Phys. Rev. Lett. \textbf{90}, 160401 (2003).}
\bibitem{Deleglise2008}{S. Del\'{e}glise, I. Dotsenko, C. Sayrin, J. Bernu, M. Brune, J-M. Raimond and S. Haroche,  Nature (London), \textbf{455}, 510 (2008).}
\bibitem{Steck2001}{D. A. Steck, W. H.  Oskay, and M. G. Raizen, Science \textbf{293}, 274 (2001).}
\bibitem{Hensinger2001}{W. K. Hensinger, H. H\"{a}ffner, A. Browaeys, N. R. Heckenberg, K. Helmerson, C. McKenzie, G. J. Milburn, W. D. Phillips, S. L. Rolston, H. Rubinsztein-Dunlop, and B. Upcroft, Nature (London) \textbf{412}, 52 (2001).}
\bibitem{Steck2002}{D. A. Steck, W. H. Oskay, and M. G. Raizen, Phys. Rev. Lett. \textbf{88}, 120406 (2002).}
\bibitem{Chaudhury2009}{S. Chaudhury, A. Smith, B. E. Anderson, S. Ghose, and P. S. Jessen, Nature (London), \textbf{461}, 768 (2009).}
\bibitem{Patil2015}{Y. S. Patil, S. Chakram, and M. Vengalattore, Phys. Rev. Lett. \textbf{115}, 140402 (2015).}
\bibitem{Java86}{J. Javanainen, Phys. Rev. Lett. \textbf{57}, 3164 (1986).}



\bibitem{Albiez2005}{M. Albiez, R. Gati, J. F\"{o}lling, S. Hunsmann, M. Cristiani, and M. K. Oberthaler, Phys. Rev. Lett. \textbf{95}, 010402 (2005).}
\bibitem{Schumm2005}{T. Schumm, S. Hofferberth, L. M. Andersson, S. Wildermuth, S. Groth, I. Bar-Joseph, J. Schmiedmayer, and P. Kr\"{u}ger, Nature Phys. \textbf{1}, 57 (2005).}
\bibitem{Levy2007}{S. Levy, E. Lahoud, I. Shomroni, and J. Steinhauer, Nature \textbf{449}, 579 (2007).}
\bibitem{Zibold10}{T. Zibold, E. Nicklas, C. Gross, and M. K. Oberthaler,
Phys. Rev. Lett. \textbf{105}, 204101 (2010).}
\bibitem{Leblanc11}{L. J. LeBlanc, A. B. Bardon, J. McKeever, M. H. T. Extavour, D. Jervis, J. H. Thywissen, F. Piazza, and A. Smerzi, Phys. Rev. Lett. \textbf{106}, 025302 (2011).}
\bibitem{Trenkwalder16}{A. Trenkwalder, G. Spagnolli, G. Semeghini, S. Coop, M. Landini, P. Castilho, L. Pezz\`{e}, G. Modugno, M. Inguscio, A. Smerzi and M. Fattori, Nat. Phys. (2016) \textbf{12}, 826 (2016).}
\bibitem{Milburn1997}{G. J. Milburn, J. Corney, E. M. Wright, D. F. Walls, Phys. Rev. A \textbf{55}, 4318 (1997).}
\bibitem{Smerzi1997}{A. Smerzi, S. Fantoni, S. Giovanazzi, and S. R. Shenoy, Phys. Rev. Lett. \textbf{79}, 4950 (1997).}
\bibitem{Raghavan1999}{S. Raghavan, A. Smerzi, S. Fantoni, and S. R. Shenoy, Phys.
Rev. A \textbf{59}, 620 (1999).}
\bibitem{Pitaevskii2001}{L. Pitaevskii and S. Stringari, Phys. Rev. Lett. \textbf{87}, 180402 (2001).}

\bibitem{Orzel2001}{C. Orzel, A. K. Tuchman, M. L. Fenselau, M. Yasuda, M. A. Kasevich, Science \textbf{291}, 2386 (2001).}
\bibitem{Esteve2008}{J. Est\`{e}ve, C. Gross, A. Weller, S. Giovanazzi, and M. K. Oberthaler, Nature (London) \textbf{455}, 1216 (2008). }
\bibitem{Gross2010}{C. Gross, T. Zibold, E. Nicklas, J. Est\`{e}ve, and M. K. Oberthaler, Nature (London) \textbf{464}, 1165 (2010).}
\bibitem{Strobel2014}{H. Strobel, W. Muessel, D. Linnemann, T. Zibold, D. B. Hume, Luca Pezz\`{e}, A. Smerzi, and M. K. Oberthaler, Science \textbf{345}, 424 (2014).}
\bibitem{Muessel2014}{W. Muessel, H. Strobel, D. Linnemann, D. B. Hume, and M. K. Oberthaler,
Phys. Rev. Lett. \textbf{113}, 103004 (2014).}

\bibitem{Britton12}{J. W. Britton, B. C. Sawyer, A. C. Keith, C.-C. J. Wang, J. K. Freericks, H. Uys, M. J. Biercuk, and J. J. Bollinger, Nature \textbf{484}, 489 (2012).}

\bibitem{Jurcevic14}{P. Jurcevic, B. P. Lanyon, P. Hauke, C. Hempel, P. Zoller, R. Blatt, and C. F. Roos, Nature \textbf{511}, 202 (2014).}

\bibitem{Richerme14}{P. Richerme,  Z.-X. Gong, A. Lee, C. Senko, J. Smith, M. Foss-Feig, S. Michalakis, A. V. Gorshkov, and C. Monroe, Nature \textbf{511}, 198 (2014).}

\bibitem{Bohnet16}{J. G. Bohnet, B. C. Sawyer, J. W. Britton, M. L. Wall, A. M. Rey, M. Foss-Feig, and J. J. Bollinger, Science \textbf{352}, 1297 (2016).}

\bibitem{Safavi18}{A. Safavi-Naini, R. J. Lewis-Swan, J. G. Bohnet, M. G\"{a}rttner, K. A. Gilmore, J. E. Jordan, J. Cohn, J. K. Freericks, A. M. Rey, and J. J. Bollinger, Phys. Rev. Lett. \textbf{121}, 040503 (2018).}

\bibitem{Everitt2009}{M. J. Everitt,  New J. Phys. \textbf{11} 013014 (2009).}

\bibitem{Devoret2013}{M.H. Devoret and R.J. Schoelkopf, Science \textbf{339} 1169 (2013).}

\bibitem{Leggett01}{A. J. Leggett, Rev. Mod. Phys. \textbf{70}, 307 (2001).}

\bibitem{Andrews97}{M. R. Andrews, C. G. Townsend, H.-J. Miesner, D. S. Durfee,
	D. M. Kurn, W. Ketterle, Science \textbf{275}, 637 (1997).}
\bibitem{Cirac1996}{J. I. Cirac, C. W. Gardiner, M. Naraschewski, and P. Zoller, Phys. Rev. A \textbf{54} R3714 (1996).}
\bibitem{Castin1997}{Y. Castin and J. Dalibard, Phys. Rev. A \textbf{55} 4330 (1997).}
\bibitem{Sinatra98}{A. Sinatra and Y. Castin, Eur. Phys. J. D \textbf{4}, 247 (1998).}


\bibitem{Hofferberth07}{S. Hofferberth, I. Lesanovsky, B. Fischer, T. Schumm and J. Schmiedmayer, Nature \textbf{449}, 324 (2007).}
\bibitem{Gring2012}{M. Gring, M. Kuhnert, T. Langen, T. Kitagawa, B. Rauer, M. Schreitl, I. Mazets, D. Adu Smith, E. Demler, and J. Schmiedmayer, Science \textbf{337}, 1318 (2012).}
\bibitem{AduSmith2013}{D. Adu Smith, M. Gring, T Langen, M. Kuhnert, B. Rauer, R. Geiger, T. Kitagawa, I. Mazets, E. Demler and J. Schmiedmayer, New J. Phys. \textbf{15}, 075011 (2013).}
\bibitem{Langen2013}{T. Langen, R. Geiger, M. Kuhnert, B. Rauer, and J. Schmiedmayer, Nature Phys. \textbf{9}, 640 (2013).}


\bibitem{Saba2005}{M. Saba, T. A. Pasquini, C. Sanner, Y. Shin, W. Ketterle, and D. E. Pritchard, Science \textbf{307}, 1945 (2005).}

\bibitem{Ruostekoski1998}{J. Ruostekoski and D. F. Walls, Phys. Rev. A \textbf{58}, R50 (1998).}
\bibitem{Huang06}{Y. P. Huang and M. G. Moore, Phys. Rev. A \textbf{73}, 023606 (2006).}
\bibitem{Khodorkovsky08}{Y. Khodorkovsky, G. Kurizki, and A. Vardi, Phys. Rev. Lett. \textbf{100}, 220403 (2008).}
\bibitem{Witthaut09}{D. Witthaut, F. Trimborn, and S. Wimberger, Phys. Rev. A \textbf{79}, 033621 (2009).}
\bibitem{Ferrini2010}{G. Ferrini, D. Spehner, A. Minguzzi, and F. W. J. Hekking, Phys. Rev. A \textbf{82}, 033621 (2010).}
\bibitem{Javanainen2013}{J. Javanainen and J. Ruostekoski, New J. Phys. \textbf{15}, 013005 (2013).}

\bibitem{Zapata2003}{I. Zapata, F. Sols, and A. J. Leggett, Phys. Rev. A 67, 021603(R) (2003).}
\bibitem{Xiong2006}{H. Xiong, S. Liu, and M. Zhan, Phys. Rev. B 73, 224505 (2006).}
\bibitem{Trujillo2009}{M. T. Martinez, A. Posazhennikova, and J. Kroha, Phys. Rev. Lett. 103, 105302 (2009).}
\bibitem{Veksler2015}{H. Veksler and S. Fishman, New J. Phys. \textbf{17},  053030 (2015).}
\bibitem{Odell2012}{D. H. J. O'Dell, Phys. Rev. Lett. \textbf{109}, 150406 (2012).}
\bibitem{Mumford2017}{J. Mumford, W. Kirkby and D. H. J. O'Dell, J. Phys. B: At. Mol. Opt. Phys. \textbf{50}, 044005 (2017).}
\bibitem{Mumford2019}{J. Mumford, E. Turner, D. W. L. Sprung and D. H. J. O'Dell, Phys. Rev. Lett. \textbf{122}, 170402  (2019).}
\bibitem{Leonhardt2002}{U. Leonhardt, Nature \textbf{415}, 406 (2002).}
\bibitem{Berry2004}{M. V. Berry  and M. R. Dennis, J. Opt. A: Pure
    Appl. Opt. \textbf{6}, S178 (2004).}
\bibitem{Berry2008}{M. V. Berry,  Nonlinearity \textbf{21}, T19 (2008).}

\bibitem{Naghiloo17}{M. Naghiloo, D. Tan, P. M. Harrington, P. Lewalle, A. N. Jordan, and K. W. Murch, Phys. Rev. A \textbf{96}, 053807 (2017).}


\bibitem{Berry1981}{M. Berry, \textit{Singularities in Waves and Rays}
    in Les Houches, Session XXXV, 1980 \textit{Physics of Defects},
    edited by R. Balian et al. (North Holland Publishing, Amsterdam,
    1981).} 

\bibitem{Thom1975}{R. Thom, \textit{Structural Stability and Morphogenesis} (Benjamin, Reading, MA, 1975).}
\bibitem{Arnold1975}{V.I. ArnolÕd, Russ. Math. Survs. 30, 1 (1975).}

\bibitem{Hohmann10}{R. H\"{o}hmann, U. Kuhl, H.-J. St\"{o}ckmann, L. Kaplan, and E. J. Heller, Phys. Rev. Lett. \textbf{104}, 093901 (2010).}   
\bibitem{Berry18}{M. V. Berry, New J.
Phys. \textbf{20}, 053066 (2018).}
\bibitem{Rooijakkers03}{W. Rooijakkers, S. Wu, P. Striehl, M. Vengalattore, and M. Prentiss, Phys. Rev. A \textbf{68}, 063412 (2003).} 
\bibitem{Huckans09}{J. H. Huckans, I. B. Spielman, B. L. Tolra, W. D. Phillips,
and J. V. Porto, Phys. Rev. A \textbf{80}, 043609 (2009).}    
\bibitem{Rosenblum14}{S. Rosenblum, O. Bechler, I. Shomroni, R. Kaner, T. Arusi-Parpar, O. Raz, and B. Dayan, Phys. Rev. Lett. \textbf{112}, 120403 (2014).}
\bibitem{Vardi2001}{A. Vardi and J. R. Anglin, Phys. Rev. Lett. \textbf{86}, 568 (2001).}

\bibitem{Paraoanu2001}{G.-S. Paraoanu, S. Kohler, F. Sols and A. J. Leggett, J. Phys. B: At. Mol. Opt.
Phys. \textbf{34}, 4689 (2001). }

\bibitem{BerryODell1999}{M. V. Berry and D. H. J. O'Dell, J. Phys. A: Math. and Gen. \textbf{32}, 3571 (1999).}

\bibitem{Sinatra2002}{A. Sinatra, C. Lobo, and Y. Castin, J. Phys. B \textbf{35}, 3599 (2002).}
\bibitem{Polkovnikov2003}{A. Polkovnikov, Phys. Rev. A \textbf{68}, 053604 (2003).}
\bibitem{Sakurai2011}{J. J. Sakurai and J. Napolitano, \textit{Modern Quantum Mechanics} 2nd Edition (Addison-Wesley, San Francisco, 2011).}
\bibitem{Hines2003}{A. P. Hines, R. H. McKenzie, and G. J. Milburn, Phys. Rev. A \textbf{67}, 013609 (2003).}


\bibitem{Lipkin65}{H.J. Lipkin, N. Meshkov, and A. J. Glick, Nucl. Phys. \textbf{62},
188 (1965).}

\bibitem{NIST}{See, M. V. Berry and C. J. Howls in \textit{NIST Digital Library of Mathematical Functions} http://dlmf.nist.gov/, Release 1.0.16 of 2017-09-18. F. W. J. Olver, A. B. Olde Daalhuis, D. W. Lozier, B. I. Schneider, R. F. Boisvert, C. W. Clark, B. R. Miller, and B. V. Saunders, eds. }

\bibitem{Krahn2009}{G. J. Krahn and D. H. J. OÕDell, J. Phys. B: At. Mol. Opt. Phys. \textbf{42}, 205501 (2009).}

\bibitem{Fadel2018}{M. Fadel, T. Zibold, B. D\'{e}camps, and P. Treutlein, Science \textbf{360}, 409 (2018).}

\bibitem{Kunkel2018}{P. Kunkel, M. Pr\"{u}fer, H. Strobel, D. Linnemann, A. Fr\"{o}lian, T. Gasenzer, M. G\"{a}rttner, and M. K. Oberthaler, Science \textbf{360}, 413 (2018).}

\bibitem{Lange2018}{K. Lange, J. Peise, B. L\"{u}cke, I. Kruse, G. Vitagliano, I. Apellaniz, M. Kleinmann, G. T\'{o}th, and C. Klempt, Science \textbf{360}, 416 (2018).}


\bibitem{Corney1998}{J. F. Corney and G. J. Milburn, Phys. Rev. A \textbf{58}, 2399 (1998).}




\bibitem{Lindblad1976}{G. Lindblad, Comm. Math. Phys. \textbf{48}, 119 (1976). }
\bibitem{Gorini1978}{V. Gorini, A. Frigerio, M. Verri, A. Kossakowski, and E. Sudarshan, Rep.  Math. Phys. \textbf{13}, 149 (1978).}
\bibitem{Bradley1997}{C. C. Bradley, C. A. Sackett, and R. G. Hulet, Phys. Rev. Lett. \textbf{78}, 985 (1997).}
\bibitem{Andrews1996}{M. R. Andrews, M.-O. Mewes, N. van Druten, D. Durfee, D. Kurn, and W. Ketterle, Science \textbf{273}, 84 (1996).}
\bibitem{Andrews1997}{M. R. Andrews, D. M. Kurn, H.-J. Miesner, D. S. Durfee, C. G. Townsend, S. Inouye, and W. Ketterle, Phys. Rev. Lett. \textbf{79}, 553 (1997).}

\bibitem{Gardiner85}{C. W. Gardiner and M. J. Collett, Phys. Rev. A \textbf{31}, 3761 (1985).}

\bibitem{Dalibard92}{J. Dalibard, Y. Castin, and K. M{\o}lmer, Phys. Rev. Lett. \textbf{68}, 580 (1992).}

\bibitem{GisinPercival1992}{N. Gisin and I. C. Percival, J. Phys. A: Math. and Gen. \textbf{25}, 5677 (1992).}

\bibitem{Carmichael93}{H. J. Carmichael, \textit{An Open System Approach to Quantum Optics}, Lecture Notes in Physics (Springer, Berlin,1993). }

\bibitem{Kosloff2013}{R. Kosloff, Entropy \textbf{15}, 2100 (2013).}
\bibitem{Gagen93}{M. J. Gagen, H. M. Wiseman and G. J. Milburn, Phys. Rev. A \textbf{48}, 132 (1993).}

\bibitem{Berry1972}{M. V. Berry and K. E. Mount, Reps. Prog. Phys \textbf{35}, 315 (1972).}

\bibitem{Nagy2011}{D. Nagy, G. Szirmai, and P. Domokos, Phys. Rev. A \textbf{84}, 043637 (2011).}

\bibitem{Bhaseen2012}{M. J. Bhaseen, J. Mayoh, B. D. Simons, and J. Keeling, Phys. Rev. A \textbf{85}, 013817 (2012).}

\bibitem{Mumford2015}{J. Mumford, D. H. J. OÕDell, and J. Larson, Ann. Phys. (Berlin) \textbf{527}, 115 (2015).}

\bibitem{Brennecke2013}{Brennecke, R. Mottl, K. Baumann, R. Landig, T. Donner, and T. Esslinger, Proc. Nat. Acad. Sci. \textbf{110},11763 (2013).}

\bibitem{Breuer_and_Pettrucione}{H.-P. Breuer and F. Pettrucione, \textit{The Theory of Open Quantum Systems} (Oxford University Press, Oxford, 2007).}

\bibitem{Ashida19}{Y. Ashida, T. Shi, R. Schmidt, H. R. Sadeghpour, J. Ignacio Cirac, and E. Demler
Phys. Rev. Lett. \textbf{123}, 183001 (2019).}

\bibitem{Das06}{A. Das, K. Sengupta, D. Sen, and B. K. Chakrabarti, Phys. Rev. B \textbf{74}, 144423 (2006).}

\bibitem{Garraway11}{B. M. Garraway, Phil. Trans. R. Soc. A \textbf{369}, 1137 (2011).}

\bibitem{buonsante12}{P. Buonsante, R. Burioni, E. Vescovi, and A. Vezzani, Phys. Rev. A \textbf{85}, 043625 (2012).}

\bibitem{lambert04}{N. Lambert, C. Emary, and T. Brandes, Phys. Rev. Lett. \textbf{92} 073602 (2004); C. Emary and T. Brandes, Phys. Rev. Lett. \textbf{90} 044101 (2003); C. Emary and T. Brandes, Phys. Rev. E \textbf{67}, 066203 (2003).}

\bibitem{mumford14a}{J. Mumford, J. Larson, and D. H. J. O'Dell, Phys. Rev. A \textbf{89}, 023620 (2014).}

\bibitem{mumford14b}{J. Mumford and D. H. J. O'Dell, Phys. Rev. A \textbf{90}, 063617 (2014).}

\bibitem{Arnold89}{V. I. Arnol'd, Mathematical Methods of Classical Mechanics, Second Edition (Springer-Verlag, New York, 1989).}

\bibitem{Choi19}{S. Choi, C. J. Turner, H. Pichler, W. W. Ho, A. A. Michailidis, Z. Papi\'{c}, M. Serbyn, M. D. Lukin, and D. A. Abanin
Phys. Rev. Lett. \textbf{122}, 220603 (2019).}

\end{thebibliography}
\end{document}